\documentclass[a4paper,12pt]{article}

\usepackage{amsmath,amssymb,amsfonts}
\usepackage{epsfig,cancel,cite}
\usepackage{dsfont}
\usepackage[usenames,dvipsnames]{color}

\newlength{\xtrawidth}
\newlength{\xtraheight}
\def\fnote#1#2{\begingroup\def\thefootnote{#1}\footnote{#2}
     \addtocounter{footnote}{-1}\endgroup}

\newcommand{\setall}{\setcounter{equation}{0}}

\begin{document}

\title{{\Large\bf Bundles over Nearly-Kahler Homogeneous Spaces in Heterotic String Theory}}
\author{Michael Klaput, Andre Lukas and Cyril Matti}
\date{}
\maketitle
\begin{center}
{\small {\it Rudolf Peierls Center for Theoretical Physics, Oxford University,\\
$~~~~~$ 1 Keble Road, Oxford, OX1 3NP, U.K.}\\
\fnote{}{m.klaput1@physics.ox.ac.uk}
\fnote{}{lukas@physics.ox.ac.uk}
\fnote{}{c.matti1@physics.ox.ac.uk}
}
\end{center}

\abstract{\noindent
We construct heterotic vacua based on six-dimensional nearly-Kahler homogeneous manifolds and non-trivial vector bundles thereon. Our examples are based on three specific group coset spaces. It is shown how to construct line bundles over these spaces, compute their properties and build up vector bundles consistent with supersymmetry and anomaly cancelation. It turns out that the most interesting coset  is $SU(3)/U(1)^2$. This space supports a large number of vector bundles which lead to consistent heterotic vacua, some of them with three chiral families.}

\newpage

\tableofcontents

%%%%%%%%%%%%%%%%%%%%%%%%%%%%%%%%%%%%%%%%%%%%%%%%%%%%%%%%%%%%%%%%%%%%%%%%%%%%%%%%%%%%%%%%%%%%%%%%%%%%%%%%%%%%%%%%%%%%%%%%%%%%%%%%%%%%%%%%%%%%%%%%
\section{Introduction}
\setall
The presence of gauge bundles is one of the distinctive features of heterotic string compactifications which is responsible for many of the physically interesting properties as well as technical complications of heterotic models. For heterotic Calabi-Yau compactifications the internal metric is not known explicitly which makes it difficult to solve for gauge connections. This problem can be largely circumnavigated by using techniques from algebraic geometry.

In this paper, we would like to study gauge bundles in the context of heterotic non-Calabi-Yau compactifications. We will construct gauge bundles leading to consistent heterotic vacua over non-Calabi-Yau spaces and we believe this to be the first time such constructions have being carried out explicitly. 

Non-Calabi-Yau compactifications, in the present context on six-dimensional manifolds with $SU(3)$ structure, are an important generalization of the Calabi-Yau case and, due to the presence of flux or torsion, they may be particularly relevant to the problem of moduli stabilisation. The conditions on the most general heterotic compactifications with a four-dimensional maximally symmetric space-time and four preserved supercharges have been found some time ago by Strominger~\cite{Strominger:1986uh}. The internal, six-dimensional manifolds of these solutions belong to a particular sub-class of manifolds with $SU(3)$ structure and they are complex but not, in general, K\"ahler. The present paper is based on a more general class of heterotic vacua which preserves only two supercharges, with the four-dimensional space-time being a domain wall~\cite{us} rather than a maximally symmetric space. Such vacua can still be associated with a covariant, four-dimensional $N=1$ supergravity theory with a  ``ground state" which is given by a half BPS domain wall. Eventually, this domain wall ground state will have to be ``lifted" to Minkowski or de-Sitter space, possibly in a way similar to what is common practice when lifting AdS vacua. 

In this paper, we work within the context of the simplest class of such heterotic domain wall vacua, where the NS flux vanishes and the dilaton is constant. In this case, it turns out that the internal six-dimensional manifold has an $SU(3)$ structure which is half-flat. Such half-flat manifolds also arise in the context of type II mirror symmetry with flux~\cite{Gurrieri:2002wz} and many useful properties of such half-flat mirror manifolds can be inferred from mirror symmetry~\cite{Gurrieri:2002wz}. Heterotic compactifications on such half-flat mirror manifolds have first been studied in Refs.~\cite{Gurrieri:2007jg, Gurrieri:2004dt} and much can be said about the gravitational sector of the resulting low-energy theories. However, the study of gauge bundles over such spaces is considerably more difficult mainly due to the lack of explicit examples. 

In this paper, we consider three explicit half-flat manifolds, the three coset spaces $SU(3)/U(1)^2$, $Sp(2)/SU(2)\times U(1)$ and $G_2/SU(3)$. In Refs.~\cite{Nolle:2010nn, Lechtenfeld:2010dr, Chatzistavrakidis:2008ii, Chatzistavrakidis:2009mh} heterotic compactifications on these coset spaces have been studied,  focusing on the gravitational sector of the theory. The main point of this paper is to study gauge bundles over these coset spaces in order to construct consistent heterotic compactifications. While these coset spaces (with their half-flat $SU(3)$ structure) are not complex and, hence, methods of algebraic geometry are difficult to apply, their group origin facilitates explicit computations. Metrics and gauge connections can be explicitly constructed and the relevant equations of 10-dimensional $N=1$ supergravity can be checked directly. We also observe that the three coset spaces have the structure of half-flat mirror manifolds so that the earlier, general results on the gravitational sector~\cite{us} directly apply. 

The plan of the paper is as follows. After a brief review of heterotic domain wall solutions in the next section, Section 3 presents the necessary material on coset spaces and the three particular examples studied in this paper. Section 4 explains how to construct vector bundles on coset spaces and how to compute some of their properties. In section 5, these general methods are then applied to our three specific coset spaces, leading up to a preliminary discussion of the model building options which arise. We conclude in Section 6. Appendix A summarises our index conventions and reviews some useful material on $SU(3)$ structures which is used in the main text. Appendix B presents details of the three coset spaces, in particular explicit generators for the Lie-algebra, structure constants and some topological invariants, such as Betti numbers.

%%%%%%%%%%%%%%%%%%%%%%%%%%%%%%%%%%%%%%%%%%%%%%%%%%%%%%%%%%%%%%%%%%%%%%%%%%%%%%%%%%%%%%%%%%%%%%%%%%%%%%%%%%%%%%%%%%%%%%%%%%%%%%%%%%%%%%%%%%%%%%%%
\section{Heterotic domain wall solutions}
\setall

%%%
Before we present the explicit coset constructions central to this paper we would like to discuss the general context of solutions to heterotic string theory into which these constructions fit. We begin with a brief review of 10-dimensional $N=1$ supergravity and its Killing spinor equations. Half-flat manifolds, of which our cosets are examples, lead to solutions of the heterotic string provided they are combined with four-dimensional domain wall solutions. In practice this means, the four-dimensional effective theory, associated to compactifications on such half-flat manifolds, is a four-dimensional covariant $N=1$ supergravity, however with a perturbative ``vacuum" solution, given by a domain wall. The structure of the relevant 10-dimensional solutions, combining half-flat manifolds and four-dimensional domain walls is explained in the second part of this introductory section. In the last part, we review the properties of a specific sub-class of half-flat manifolds, so called half-flat mirror manifolds, which arise in the context of type II mirror symmetry with NS flux~\cite{Gurrieri:2002wz}. As we will show in the next section, our coset manifolds fall into this sub-class of half-flat mirror manifolds.
%%%%%%
\subsection{Ten-dimensional $N=1$ supergravity}
The bosonic field content of ten-dimensional $N=1$ supergravity consists of the metric $\hat G$, the dilaton $\hat\phi$, the NS-NS two-form $\hat B$  with associated field strength $\hat H=d\hat B$ and the gauge field $\hat A$ with gauge group $E_8\times E_8$ or $SO(32)$ and field strength
\begin{equation}
  \hat F=d\hat A+\hat A\wedge\hat A \;.
\end{equation}
Much of the discussion in this paper applies to both gauge groups, however, when we need to be specific we will focus on the $E_8\times E_8$ case. 
%%%
The bosonic part of the action governing the dynamics of these fields is given by
\begin{equation}\label{action}
	S=-\frac{1}{2\kappa^2_{10}}\int_{M_{10}} e^{-2\hat\phi}\left[\hat{R}*\textbf{1}-4d\hat\phi\wedge *d\hat\phi+\frac{1}{2}\hat H\wedge *\hat H+\frac{\alpha'}{4} e^{\hat\phi}\left({\rm tr}\hat F\wedge *\hat F-{\rm tr} \hat{R}\wedge*\hat R\right)\right]\;,
\end{equation}
to first order in the string tension $\alpha'$. Here $\kappa_{10}$ is the 10-dimensional Planck constant. This action has to be supplemented by the Bianchi identity
\begin{equation}\label{bianchi}
  d\hat H=\frac{\alpha'}{4}\left({\rm tr}\hat R\wedge \hat R-{\rm tr}\hat F\wedge \hat F\right).
\end{equation}
%%%
The fermionic field content consists of the gravitino, $\psi_M$, the dilatino, $\lambda$ and the gauginos $\chi$, all of them 10-dimensional Majorana-Weyl spinors. Their supersymmetry transformations take the form
\begin{subequations}
\begin{align}
	\delta\psi_M&=\left(\nabla_M+\frac{1}{8}{\cal\hat H}_M\right)\epsilon\;,\label{dpsi}\\
	\delta\lambda&=\left(\not\!\nabla\hat\phi+\frac{1}{12}{\cal\hat H}\right)\epsilon\;,\label{dlambda}\\
	\delta\chi&=\hat F_{MN}\Gamma^{MN}\epsilon\; ,\label{dchi}
\end{align}	
\end{subequations}
where $\epsilon$ is a 10-dimension Majorana-Weyl spinor parametrizing supersymmetry. Further, we have introduced the gamma matrix contractions ${\cal\hat H}=\hat H_{MNP}\Gamma^{MNP}$ and ${\cal\hat H}_M=\hat H_{MNP}\Gamma^{NP}$.

Supersymmetric solutions of the theory should satisfy the Killing spinor equations $\delta\psi_M=0$, $\delta\lambda=0$ and $\delta\chi=0$. It then follows that they satisfy the equations of motion derived from the action~\eqref{action}, provided that, in addition, the Bianchi identity~\eqref{bianchi} is satisfied. We will now introduce the particular general class of solutions relevant for this paper.

%%%%%%
\subsection{Domain walls and SU(3)-structures}\label{section_domainwalls}
The ``traditional" approach to finding 10-dimensional solutions to the heterotic string which may lead to phenomenologically interesting compactifications to four dimensions is based on assuming four preserved supercharges and an external four-dimensional space-time which is maximally symmetric. In the simplest case, that is, for vanishing $H$-flux and a constant dilaton, this approach leads to internal Calabi-Yau manifolds times a four-dimensional Minkowski space-time~\cite{Candelas:1985en}. These solutions have been the basis of much research and attempts to relate the heterotic string to observable physics. More generally, one can allow for non-vanishing flux and a varying dilaton but keep the requirement of four preserved supercharges and maximally-symmetric external space-time. This leads to a set of solutions, based on complex, non-K\"ahler manifolds, described by Strominger~\cite{Strominger:1986uh}. Unfortunately, not many examples of such manifolds are known explicitly. 

Here, we will take a somewhat different approach. We will only ask for two preserved supercharges and allow the four-dimensional space-time to be a domain wall, rather than a maximally symmetric space. The detailed implications of this approach and the structure of the solutions has been worked out in Ref.~\cite{us}. In particular, it was found that, in the simplest case for vanishing flux and a constant dilaton, the internal space is now an $SU(3)$ structure manifold of a particular kind, namely a so-called half-flat manifold. How can 10-dimensional solutions with two supercharges and based on a four-dimensional domain wall be phenomenologically relevant? As is known for some time~\cite{Gurrieri:2007jg, Gurrieri:2004dt}, compactifications of the heterotic string on half-flat manifolds are associated to perfectly covariant four-dimensional $N=1$ supergravity theories. However, due to a non-trivial superpotential, the ``vacuum state" of these four-dimensional supergravities is not a maximally symmetric space but a domain wall, precisely the same domain wall which appears in the full 10-dimensional solution. 
While such a domain wall state is ultimately not a desired ground state for our four-dimensional universe it might be argued that ``lifting" it to, say, a Minkowski or de Sitter space, by means of additional contributions to the scalar potential, is no more a far-fetched scenario than the widespread practice of lifting an AdS vacua. At any rate, this is the philosophy behind our approach and the motivation to look at more detailed, phenomenologically relevant properties of such domain wall vacua. 

Let us now discuss the structure of these solutions in more details. As mentioned, we will do this for the simplest case of vanishing NS flux and constant dilaton, that is,
\begin{equation}
 \hat{H}=0\; ,\quad \hat{\phi}=\mbox{constant}\; . \label{Hphi}
\end{equation}
The metric, consisting of a six-dimensional internal space and a four-dimensional domain wall, has the structure
\begin{equation}\label{metric}
	ds_{10}^2=\eta_{\alpha\beta} dx^\alpha dx^\beta+dy^2+g_{uv}(x^m)dx^udx^v\; .
\end{equation}
Here the indices $\alpha,\beta,\dots$ range over $0,1,2$ and label the world volume coordinates of the domain wall, while $y=x^3$ is the remaining direction of four-dimensional space-time transverse to the domain wall. The six directions labeled by indices $u,v,\dots=4,...,9$ refer to the internal, compact manifold, $X$, while indices $m,n,\dots =3,\ldots,9$ label all seven directions transverse to the domain wall. More details on our conventions can be found in Appendix~\ref{Conventions}. Given the absence of stress energy due to Eqs.~\eqref{Hphi} the above metric needs to be Ricci-flat in order to solve Einstein's equations. This means that the seven-dimensional space transverse to the domain wall must be a manifold with holonomy $G_2$ (or smaller), carrying a covariantly constant spinor $\eta=\eta (x^m)$. The Killing spinor equations associated to the supersymmetry transformations~\eqref{dpsi} and \eqref{dlambda} of the gravitino and dilatino are then satisfied for 10-dimensional spinors of the form
\begin{equation}
 \epsilon(x^m)=\rho\otimes \eta(x^m)\otimes\theta\; ,\label{spinoransatz}
\end{equation}
where $\rho$ is a $2+1$--dimensional spinor on the domain wall world volume which parameterises the two supersymmetries of the solution. The appearance of $\theta$ is due to the dimensionality of the respective spinors and is a constant two-component spinor depending on the chirality of $\epsilon$ and the choice of gamma matrices representation. Manifolds with $G_2$ holonomy can also be characterised by a torsion-free $G_2$ structure, that is, a closed and co-closed three form $\varphi$. In terms of the covariantly constant spinor $\eta$, this form can be written as $\varphi_{mnp}=-i\eta^\dagger\gamma_{mnp}\eta$. In order to understand the structure of the six-dimensional internal space $X$ is it useful to decompose the $G_2$ structure in the usual way as
\begin{equation}
 \varphi=dy\wedge J+\Omega_-\; ,\quad \star\varphi =dy\wedge\Omega_++\frac{1}{2}J\wedge J\; ,
\end{equation}
where $J$ and $\Omega=\Omega_++i\Omega_-$ are two- and three-forms, respectively, which define an $SU(3)$ structure on $X$ with associated metric $g_{uv}$.  Closure and co-closure of $\varphi$ then translates into the conditions
\begin{equation} 
\begin{array}{lllrrrl}
	d\Omega_-&=&0&,&	J\wedge\ dJ&=&0\\
	d\Omega_+&=&J\wedge\partial_{y} J&,&	dJ&=&\partial_{y}\Omega_-.
\end{array}\label{HF1}
\end{equation}
The first two of these equations tell us that the $SU(3)$ structure on $X$ has specific properties which are referred to as ``half-flat". In terms of the classification of $SU(3)$ structures by five torsion classes $W_1,\ldots , W_5$ a half-flat manifold can be characterised by
\begin{equation}\label{torsion}
W_{1-}=W_{2-}=W_4=W_5=0\; ,
\end{equation}
with the remaining classes being arbitrary. A short summary of $SU(3)$ structures and torsion classes is presented in Appendix~\ref{appendix_su3structure}. We note that, since $W_1$ and $W_2$ are generically non-vanishing, the almost complex structure $J$ is not integrable, so that half-flat manifolds are, in general, not complex. The last two equations~\eqref{HF1} describe how the $SU(3)$ structure on $X$ varies along the direction $y$ and are known as Hitchin's flow equations~\cite{Lopes Cardoso:2002hd, Hitchin}. To summarise the discussion so far, we have introduced a class of space-time background solutions to the heterotic string with a ``dual" interpretation. From one point of view these backgrounds consist of $2+1$-dimensional Minkowski space times a manifold with $G_2$ holonomy. Alternatively, they can be viewed as a four-dimensional domain wall with transverse direction $y$ and a six-dimensional half-flat manifold fibered along this transverse direction. Of course, this should only be considered a solution at lowest, zeroth order in $\alpha'$ and, as such, it does not incorporate the gauge fields which are the characteristic feature of the heterotic string. 

The main point of this paper is to develop this class of solutions beyond zeroth order in $\alpha'$ and include non-trivial gauge fields. For simplicity we will consider purely internal gauge fields $F_{uv}$ on the six-dimensional half-flat manifold $X$, with all other components vanishing. In order for those gauge fields to preserve the two supersymmetries the gaugino supersymmetry variation~\eqref{dchi} has to vanish for spinors of the form~\eqref{spinoransatz}. This implies the constraints
\begin{equation}\label{HYM}
  \Omega\,\neg\, F =0\;,\quad J\,\neg\, F=0\; .
\end{equation}
where the symbol $\neg$ denotes contraction over two indices. In the Calabi-Yau case, the first equation~\eqref{HYM} implies that $A$ is a connection on a holomorphic vector bundle, while the second one, via the Donaldson-Uhlenbeck-Yau theorem, is equivalent to saying that this vector bundle is slope-stable with slope zero. For Calabi-Yau manifolds these statements provide the only practical way of solving the equations since neither the metric nor the connection are explicitly known. For half-flat manifolds we are not aware of the existence of analogous theorems. We will circumnavigate this problem by working with specific half-flat manifolds, to be introduced later, for which metric and gauge connections can be written down explicitly. For these, the two conditions~\eqref{HYM} can then be checked directly. 

In addition to the above equations, we also have to satisfy the Bianchi identity~\eqref{bianchi}. Since the left-hand side of this equation is exact the right-hand side needs to be trivial in cohomology. Splitting up the gauge field strength into an observable $E_8$ part $F$ and a hidden $E_8$ part $\tilde{F}$ (both of which have to satisfy~\eqref{HYM}) this leads to the integrability condition
\begin{equation}\label{biancoho}
  \left[{\rm tr}\hat R\wedge \hat R\right]= \left[{\rm tr}F\wedge F+{\rm tr}\tilde F\wedge \tilde F\right]\; ,
\end{equation}
where the square bracket denotes the cohomology class. Satisfying this condition is necessary and sufficient for a solution to the Bianchi identity. Unless the right-hand side of~\eqref{bianchi} cancels point-wise (rather than merely in cohomology) this solution will require a non-vanishing NS flux $H$ at order $\alpha'$. This NS flux will feed into the Killing spinor equations for gravitino and dilatino and generate order $\alpha'$ corrections to the metric and the dilaton. Here, we will not attempt to calculate these $\alpha'$ corrections explicitly but assume that our compactification is at sufficiently large radius for them to be small. However, we will ensure that the integrability condition~\eqref{biancoho} is satisfied for our explicit examples.
%%%%%%
\subsection{Half-flat mirror manifolds}\label{mirrorhalflat}
Before we move on to describe our specific examples we would like to introduce a special class of half-flat manifolds, so-called half-flat mirror manifolds. They have been introduced in Ref.~\cite{Gurrieri:2002wz} in order to understand type II mirror symmetry with NS flux. The relation to mirror symmetry implies some additional properties of these manifolds which have been derived in Ref.~\cite{Gurrieri:2002wz} and will be reviewed below. These properties facilitate string compacitification and an explicit calculation of the effective four-dimensional theory. In Refs.~\cite{Gurrieri:2004dt, Gurrieri:2007jg} this has been used to work out the four-dimensional theory from compactifications of the heterotic string on half-flat mirror manifolds. In the present context they are relevant because the particular half-flat manifolds used in this paper share the specific properties of half-flat mirror manifolds, as we will show. In particular, this means that the general results for the four-dimensional effective theory obtained in Refs.~\cite{Gurrieri:2004dt, Gurrieri:2007jg} apply to compacitifications on these manifolds. 

Half-flat mirror manifolds are equipped with a set, $\{\omega_\jmath\}$, of two-forms, a dual set, $\{\tilde{\omega}^\jmath\}$, of four-forms and a ``symplectic" set $\{\alpha_{\cal A},\beta^{\cal B}\}$ of three-forms, satisfying the integral relations
\begin{equation}\label{Jbasisint}
  \int \omega_\imath\wedge\tilde\omega^\jmath=\delta_\imath^\jmath\; ,\quad
  \int \alpha_{\cal A}\wedge\alpha_{\cal B}=0\; , \quad \int \beta^{\cal A}\wedge\beta^{\cal B}=0\; , \quad \int \alpha_{\cal A}\wedge\beta^{\cal B}=\delta_{\cal A}^{\cal B}\; .
\end{equation}
This is analogous to Calabi-Yau manifolds, however, unlike in the Calabi-Yau case not all of these forms are closed. Specifically, the exterior derivatives of the non-closed forms are given by
\begin{equation}
d\omega_\imath=e_\imath\beta^0\; ,\quad d\alpha_0=e_\imath\tilde{\omega}^\imath\; , \label{hfmdef}
\end{equation}
In complete analogy with Calabi-Yau manifolds, the forms $J$ and $\Omega$ which define the $SU(3)$ structure can be expanded as
\begin{equation}\label{Jexpansion}
	J=v^\imath\omega_\imath\; ,\quad\Omega={\cal Z}^{\cal A}\alpha_{\cal A}-\mathcal{G}_{\cal A}\beta^{\cal A}\; .
\end{equation}
From Eqs.~\eqref{hfmdef} these have non-vanishing exterior derivatives
\begin{equation}
	dJ=v^\imath e_\imath\beta^0\; ,\quad d\Omega={\cal Z}^0e_\imath\tilde\omega^\imath\; , 
\end{equation}
an indication that these are $SU(3)$ structure rather than $SU(3)$ holonomy manifolds. The torsion classes characterising the $SU(3)$ structure can be read of from the right-hand sides of these equations and comparison with Appendix~\ref{appendix_su3structure} shows that they indeed satisfy the characteristic half-flat constraints~\eqref{torsion}. 

%%%%%%%%%%%%%%%%%%%%%%%%%%%%%%%%%%%%%%%%%%%%%%%%%%%%%%%%%%%%%%%%%%%%%%%%%%%%%%%%%%%%%%%%%%%%%%%%%%%%%%%%%%%%%%%%%%%%%%%%%%%%%%%%%%%%%%%%%%%%%%%%
\section{Nearly-Kahler homogeneous spaces}\label{chapter3}
\setall

%%%
In this section, we introduce the particular six-dimensional manifolds on which we would like to compactify heterotic string theory. Vector bundles and gauge connections on these manifolds will be discussed in the following section. The general class of manifolds from which we would like to draw are homogeneous spaces, that is, coset spaces of Lie groups. It is known~\cite{Butru} that precisely four six-dimensional spaces within this class are half-flat manifolds, namely the cosets $SU(3)/U(1)^2$, $Sp(2)/SU(2)\times U(1)$, $G_2/SU(3)$ and $SU(2) \times SU(2)$. We will indeed see that the torsion classes of these manifolds satisfy the half-flat constraints~\eqref{torsion} and are, in fact, somewhat more special in a way that is referred to as ``nearly Kahler". In addition, we will also systematically construct an explicit set of forms on these manifolds which satisfy the relations~\eqref{hfmdef} for half-flat mirror manifolds. This has been first exposed by House and Palti for the $SU(3)/U(1)^2$ case~\cite{House:2005yc}.
To set the scene, we begin by reviewing some well-known facts on coset spaces~\cite{Castellani:1999fz,Kapetanakis:1992hf,MuellerHoissen:1987cq,Camporesi:1990wm,Castellani:1983tb,Lust:1986ix,KashaniPoor:2007tr} and $SU(3)$ structures on such spaces. This formalism is then applied to the above near-Kahler coset spaces, mainly following the results of Ref.~\cite{Chatzistavrakidis:2008ii,Chatzistavrakidis:2009mh}. Since the space $SU(2) \times SU(2)$ is less suited for bundle constructions it will not be discussed explicitly and we will focus on the first three examples.

%%%%%%
\subsection{Coset space formalism}
In the following we collect some of the required results for coset spaces~\cite{Castellani:1999fz,Kapetanakis:1992hf,MuellerHoissen:1987cq,Camporesi:1990wm,Castellani:1983tb,Lust:1986ix}, mainly to introduce the relevant notation and conventions. Let $G$ be a Lie-group and $H$ a sub Lie-group of $G$. The coset space $G/H$ is defined as the set of left cosets which arise from the equivalence relation
\begin{equation}
  g\sim g'\Leftrightarrow g^{-1}g'\in H\; .
\end{equation}
This means two elements, $g$ and $g'$ of $G$ are considered to be equivalent if they can be related by right multiplication with some element of the subgroup $H$. A useful way to think about the group $G$ in this context, which we will make use of later, is as a principal bundle $G(G/H,H)$ with base space $G/H$ and fibers given by the orbits of $H$. The Lie-algebra $\mathds{G}$ of $G$ can be written as a direct sum
\begin{equation}\label{decomp}
  \mathds{G}=\mathds{H}\oplus\mathds{K}\;,
\end{equation}
where $\mathds{H}$ is the Lie-algebra of the sub-group $H$ and $\mathds{K}$ is the remainder, which corresponds to the coset. In the following, we will adopt the following conventions
\begin{equation}
\quad T_A\in\mathds{G}, \quad H_i\in\mathds{H}, \quad K_a\in\mathds{K}
\end{equation}
 to denote Lie algebra basis elements in those various parts. Here, indices $A, B, C, \dots$ run over the whole Lie algebra $\mathds{G}$, while $a, b, c, \dots$ denote coset indices and $i, j, k, l, \dots$ label directions in $\mathds{H}$. Our conventions are summarized in Appendix~\ref{Conventions}.
 The structure constants, ${f_{AB}}^C$, are split up into different types accordingly. Coordinates relative to the basis $\{K_a, H_i\}$ are denoted as $(x^a,z^i)$. 

For reasons which will become apparent in Section \ref{section_vectorbundles} we also require the Lie Algebra $\mathds{G}$ to decomposes reductively, that is, we can choose a basis such that the structure constants satisfy
\begin{equation}\label{eq_reductivity}
f_{ia}^{\phantom{ia}j}= 0 \qquad f_{ij}^{\phantom{ij}a}=0\; .
\end{equation}
It turns out that this can indeed be achieved for all our explicit examples. For such a reductive decomposition, the non-vanishing commutation relations take the form
\begin{eqnarray}\nonumber
  \left[K_a,K_b\right]&=&f_{ab}^{\phantom{ab}c}K_c+f_{ab}^{\phantom{ab}i}H_i,\\\label{structconst}
  \left[H_i,K_a\right]&=&f_{ia}^{\phantom{ia}b}K_b,\\\nonumber
  \left[H_i,H_j\right]&=&f_{ij}^{\phantom{ij}k}H_k
\;.
\end{eqnarray}
In practice, the relevant geometrical information about the coset is contained in the structure constants. For our three examples they are explicitly given in Appendix \ref{Data}.

%%%
In order to get an explicit description of the coset space one can choose one representative for each coset. Using the exponential map, such a representative can be written as
\begin{equation}\label{expo}
L(x)={\rm exp}(x^aK_a)\; .
\end{equation}
In more mathematical language, $L$ can be viewed as a section of the principal bundle $G(G/H,H)$. A non-singular set of one-forms on $G/H$ can be obtained following a procedure analogous to the one leading to left-invariant one-forms on $G$. First, define the Lie-algebra valued one-form
\begin{equation}\label{form}
V=L^{-1}dL\;,
\end{equation}
where $d$ is the exterior derivative on $G/H$. Then expand $V$ in terms of the chosen Lie-algebra basis as
\begin{equation}
V=e^aK_a+\varepsilon ^i H_i \; \label{Vdef}
\end{equation}
with one-form ``coefficients" $e^a$ and $\epsilon^i$. It can be shown that the one-forms $e^a$, in the directions of the coset generators $K_a$, are indeed non-singular, that is, they form a basis of the co-tangent space on $G/H$ and can be used as a vielbein. The algebra of their exterior derivatives follows from the Maurer-Cartan structure equations on $G$. Using the commutation relations~\eqref{structconst} one obtains
\begin{eqnarray}\label{d}
  de^a&=&-\frac{1}{2}f_{bc}^{\phantom{bc}a}e^b\wedge e^c-f_{ib}^{\phantom{ib}a}\varepsilon^i\wedge e^b,\\
  d\varepsilon^i&=&-\frac{1}{2}f_{ab}^{\phantom{ab}i}e^a\wedge e^b-\frac{1}{2}f_{jk}^{\phantom{jk}i}\varepsilon^j\wedge\varepsilon^k\; .
\end{eqnarray}
While the forms $e^a$ are left-invariant when viewed as forms on the group $G$ this is no longer the case when they descend to the coset $G/H$. For the subsequent discussion we will need to know the $G$-transformations of the $e^a$ explicitly, so we briefly discuss their derivation~\cite{Castellani:1999fz, Kapetanakis:1992hf, Camporesi:1990wm, MuellerHoissen:1987cq}. Suppose the left-action of an element $g\in G$ on $G/H$ maps a coset represented by $L(x)$ into a coset represented by $L(x')$. Then we can write
\begin{equation}\label{magic}
  gL(x)=L(x')h\; ,
\end{equation}
where the ``gauge transformation" with $h\in H$ on the right-hand side accounts for the fact that the group action, while leading to an element in the coset represented by $L(x')$, does not necessarily give the chosen representative $L(x')$. Using the definition~\eqref{form} of the Lie-algebra valued one-form $V$ this transformation law translates to 
\begin{equation}\label{transformer}
  V(x')=hV(x)h^{-1}+hdh^{-1}.
\end{equation}
From the coset point of view, the second term does not change the equivalence class of $V$ and can be discarded. Consequently, the basis forms $e^a$ transform in the adjoint representation $D$ of $H$ as 
\begin{equation}
  e^a(x')=D_b^{\phantom{b}a}(h^{-1})e^b(x)\; . \label{etrafo}
\end{equation}
For an infinitesimal $G$-action $g={\bf 1}+\epsilon^AT_A$, the associated gauge transformation $h$ in Eq.~\eqref{magic} can be written as $h={\bf 1}-\epsilon^A{W_A}^iH_i$, with ``compensator" functions ${W_A}^i$. Expanding the exponentials in Eq.~\eqref{magic}, these functions can be calculated order by order but their explicit form will not be needed in the present context. Inserting into Eq.~\eqref{etrafo} the infinitesimal transformation of the vielbein becomes
\begin{equation}
 e^a(x')-e^a(x)=\epsilon^AW_A^{\phantom{A}i}f_{bi}^{\phantom{bi}a}e^b(x)\;. \label{einftrafo}
\end{equation}
This transformation law will be crucial in a moment when we establish what it means for an $SU(3)$ structure on a coset space $G/H$ to be $G$-invariant.

%%%%%%
\subsection{G-invariant structures}
As discussed earlier, an $SU(3)$ structure is given by a two-form $J$ and a three-form $\Omega$ which are subject to the conditions~\eqref{comp}. Associated to such an $SU(3)$ structure is a metric $g$ which can be computed from $J$ and $\Omega$. On a coset space $G/H$ these tensors can be expanded in terms of the vielbein forms $e^a$ as
\begin{equation}\label{eq_formsdef}
J = \frac{1}{2!}J_{ab} \; e^a \wedge e^b\;, \quad
\Omega = \frac{1}{3!}\Omega_{abc}\;e^a \wedge e^b \wedge e^c\; ,\quad
g=g_{ab}\; e^a\otimes e^b\; .
\end{equation}
The construction of $SU(3)$ structures on coset spaces $G/H$ adopted in this paper is based on two additional assumptions. First, we assume that the coefficients, $J_{ab}$, $\Omega_{abc}$ and $g_{ab}$ in the above expressions are constant, rather than more general functions on the coset space. Secondly, we require that $J$, $\Omega$ and $g$ are invariant under $G$ actions on the coset. In order to work out the implications of this second assumptions we need to transform the expansions~\eqref{eq_formsdef} using the transformation law~\eqref{einftrafo} of the vielbein forms. When applied, for example, to the metric this leads to
\begin{equation}
  g_{ab}\; e^a(x')\otimes e^b(x')=g_{ab}\; e^a(x)\otimes e^b(x)+g_{ab}\;\epsilon^AW_A^i\left(f_{ci}^{\phantom{ci}a}e^c\otimes e^b+f_{di}^{\phantom{di}b}e^d\otimes e^a\right)\; .
\end{equation}
Then, $G$-invariance requires the second term of the right hand side to vanish. Hence, the coefficients $g_{ab}$ of a $G$--invariant metric are constrained by
\begin{equation}\label{ginv}
  f_{i(a}^{\phantom{i(a}c} g_{ b)c} =0 \;.
\end{equation}
The same steps can be repeated for $J$ and $\Omega$ which turn out to be $G$--invariant if their components satisfy
\begin{equation}\label{Jinv}
  f_{i[a}^{\phantom{a]i}c}J_{b]c}=0\; ,\quad
  f_{i[a}^{\phantom{a]i}d}\Omega_{bc]d}=0\; .
\end{equation}
To summarise, a $G$-invariant $SU(3)$ structure with constant coefficients on $G/H$ is given by forms $J$ and $\Omega$ as in Eq.~\eqref{eq_formsdef} with their coefficients satisfying the constraints~\eqref{Jinv}.

Of course, it is not clear that all $SU(3)$ structures or even all half-flat $SU(3)$ structures on $G/H$ are $G$-invariant and have constant coefficient expansions in terms of the forms $e^a$. So we should keep in mind that the moduli space of such structures might be larger than we will derive below. An exploration of this full moduli space is beyond the scope of the present paper. The important fact, for the purpose of this paper, is that the two additional assumptions greatly simplify the technical problems and do allow for half-flat $SU(3)$ structures on the relevant coset spaces. This means that, within the context of the domain wall solutions discussed earlier, we can indeed consider compactifying the heterotic string on these spaces.
%%%%%%
\subsection{Specific coset spaces}
We will now apply the formalism outlined previously to our three examples $SU(3)/U(1)^2$, $Sp(2)/SU(2)\times U(1)$ and $G_2/SU(3)$, largely following the discussion in Ref.~\cite{Chatzistavrakidis:2008ii,Chatzistavrakidis:2009mh}. In practice, this means finding families of $G$--invariant $SU(3)$ structures which satisfy the half-flat conditions required for our heterotic compactifications. To avoid cluttering the main text, the relevant group-theoretical information, such as generators and structure constants, has been collected in Appendix~\ref{appendix_su3structureconstants}. We will also demonstrate the existence of sets of forms on these cosets which satisfy the characteristic relations~\eqref{hfmdef} of half-flat mirror manifolds. From hereon, our convention is to have coset indices $a,b,c,\dots$ run over values $1,\ldots ,6$. Indices $i,j,k,\dots$, which label the generators of the sub-group $H$, range from $7,\ldots ,{\rm dim}(G)$. 

%%%%%%
\subsubsection{$SU(3)/U(1)^2$}
For the $SU(3)$ generators $T_A$ we choose the usual Gell-Mann matrices, however, relabeled in such a way that the coset generators, corresponding to the non-diagonal Gell-Mann matrices, carry indices from $1$ to $6$. The resulting generators and structure constants are given in appendix \ref{appendix_su3structureconstants}. Solving Eq.~\eqref{ginv} shows that the most general $SU(3)$--invariant metric takes the form
\begin{equation}
  ds^2=R_1^2\;(e^1\otimes e^1+e^2\otimes e^2)+R_2^2\;(e^3\otimes e^3+e^4\otimes e^4)+R_3^2\;(e^5\otimes e^5+e^6\otimes e^6),
  \label{su3metric}
\end{equation}
where $R_1$, $R_2$ and $R_3$ are arbitrary real parameters representing the moduli.

%%%
Moreover, solving the Eqs.~\eqref{Jinv} one finds that the space of $G$-invariant two- and three-forms is spanned by
\begin{equation}
e^{12}\; , \quad e^{34}\;, \quad e^{56}\; , \quad  e^{136}-e^{145}+e^{235}+e^{246}\; ,\quad e^{135}+e^{146}-e^{236}+e^{245}\; .
\end{equation}
Possible $G$-invariant structures $(J,\Omega)$ are linear combinations of these forms which are further restricted by having to satisfy the compatibility conditions~\eqref{comp} for $SU(3)$ structures. This lead to the most general solution
\begin{eqnarray}
  J&=&R_1^2\;e^{12}-R_2^2\;e^{34}+R_3^2\;e^{56}\label{JSU3}\\
  \Omega&=&R_1R_2R_3
\left[ 
\left( e^{136}-e^{145}+e^{235}+e^{246} \right) 
+ 
\mathrm{i}\, 
\left( e^{135}+e^{146}-e^{236}+e^{245} \right) 
\right] \;\label{OmegaSU3}
\end{eqnarray}
By computing the metrics associated to this family of $SU(3)$ structures one can verify that the moduli $R_1$, $R_2$ and $R_3$ are indeed identical to the ones appearing in \eqref{su3metric}.

We can now introduce the following linear combinations of $G$-invariant two-, three- and four-forms.
\begin{align}\label{eq_su3basisstart}
  \omega_1 &=-\frac{1}{(2\pi)}\left(e^{12}+\frac{1}{2}\, e^{34}-\frac{1}{2}\, e^{56}\right)
& 
\tilde\omega^1 &=\frac{2(2\pi)}{3\,{\cal V}}\left(2\,e^{1234}+e^{1256}-e^{3456}\right)
\\
  \omega_2 &=-\frac{1}{2(2\pi)}\left(e^{12}+e^{34}\right)
& 
\tilde\omega^2 &=-\frac{2(2\pi)}{{\cal V}}\left(e^{1234}+e^{1256}\right)\label{eq_su3end}
\\
 \omega_3 &= \frac{2}{3(2\pi)}\left(e^{12}-e^{34}+e^{56}\right)
& 
\tilde\omega^3 &=\frac{(2\pi)}{{2\,\cal V}}\left(e^{1234}-e^{1256}+e^{3456}\right)
\\
  \alpha_0 &=\frac{(2\pi)}{4{\cal V}}\left(e^{136}-e^{145}+e^{235}+e^{246}\right)
& 
\beta^0 &=\frac{1}{(2\pi)}\left(e^{135}+e^{146}-e^{236}+e^{245}\right)\;,\label{eq_su3basisend}
\end{align}
where ${\cal V}$ is the volume of the coset space defined by
\begin{equation}\label{vol}
  {\cal V}=\int_X e^{123456}\;.
\end{equation}
Introducing the volume factors in the above definitions ensures that these forms satisfy the standard integral normalisations~\eqref{Jbasisint}. Moreover, using the Maurer-Cartan relation~\eqref{d} together with the explicit structure constants, it can be verified that they satisfy the defining differential relations~\eqref{hfmdef} for half-flat mirror manifolds with intrinsic torsion parameters given by
\begin{equation}
  e_1=0\;, \quad e_2=0\;, \quad e_3=1\;. \label{SU3e}
\end{equation}
The $SU(3)$ structure forms $J$ and $\Omega$ in Eqs.~\eqref{JSU3} and \eqref{OmegaSU3} can also be expanded in terms of this new basis. This leads to expressions which conform to the general ones~\eqref{Jexpansion} for half-flat mirror manifolds. Altogether, this shows that $SU(3)/U(1)^2$ can be given the structure of a half-flat mirror manifold.

In particular, from the general discussion in Section~\ref{mirrorhalflat}, this means that the $SU(3)$ structures defined above are half-flat. 
Explicitly, the torsion classes are given by~\cite{Chatzistavrakidis:2009mh},
\begin{eqnarray}
W_1^+ &=& \frac{\left(R_1^2+R_2^2+R_3^2\right)}{3R_1R_2R_3}\;,\\
W_2^+ &=& \frac{2}{3R_1R_2R_3}\left[R_1^2\left(2R_1^2-R_2^2-R_3^2\right)e^{12}-R_2^2\left(2R_2^2-R_1^2-R_3^2\right)e^{34}\right.\nonumber\\
&&\left.+R_3^2\left(2R_3^2-R_1^2-R_2^2\right)e^{56}\right]
\end{eqnarray}
It will be relevant to note that on the locus in moduli space where the three radii are equal, $R_1=R_2=R_3\equiv R$, the torsion classes reduce to
\begin{equation}
W_1^+ =\frac{1}{R}\;,\quad W_2^+=0\; .
\end{equation}
This shows that the $SU(3)$ structure is nearly-Kahler at this particular locus.

Writing $J$ in \eqref{JSU3} in terms of the above two-forms $\omega_\imath$ and comparing with Eq.~\eqref{Jexpansion} we can
read off expressions for ``Kahler" moduli $v^\imath$, defined in the context of half-flat mirror manifolds. In terms of the radii $R_i$, they are given by
\begin{equation}
    v^1=-\frac{4\pi}{3} \left( R_1^2 +  R_2^2 -2 R_3^2\right)\; ,\quad
  v^2=4\pi\left( R_2^2 - R_3^2\right)\; ,\quad
  v^3=\pi \left( R_1^2 +  R_2^2 +R_3^2\right)\; .
\end{equation}
Note that the forms $\omega_1$ and $\omega_2$ are closed and, hence, define cohomology classes, while $\omega_3$ is not closed. We, therefore, expect two massless modes, $v^1$ and $v^2$, and one massive one, $v^3$. This expectation is confirmed by looking at the superpotential for half-flat mirror compactifications~\cite{Gurrieri:2004dt} which is given by 
\begin{equation}
W=e_\imath T^\imath\; ,\quad \mbox{where } {\rm Re}(T^\imath)=v^\imath\; . \label{W}
\end{equation}
In view of the torsion parameters~\eqref{SU3e} this means
\begin{equation}
 W=T^3\; , 
\end{equation}
so that $T^1$ and $T^2$ are indeed massless. Also note, the existence of only two $G$-invariant three-forms, $\alpha_0$, $\beta^0$ means that the analogues of complex structure moduli are not present in this particular model.

%%%%%%
\subsubsection{$Sp(2)/SU(2)\times U(1)$}
In order to obtain a half-flat space, this coset is defined by taking the non-maximal embedding of $SU(2)$ into $Sp(2)$. Group-theoretical details, in particular generators and structure constants, are again given in Appendix \ref{appendix_sp2structureconstants}. We proceed in the same way as in the previous case. Solving Eq.~\eqref{ginv} the most general $Sp(2)$--invariant metric turns out to be
\begin{equation}\label{Spmetric}
  ds^2=R_1^2\;(e^1\otimes e^1+e^2\otimes e^2)+R_2^2\;(e^3\otimes e^3+e^4\otimes e^4)+R_1^2\;(e^5\otimes e^5+e^6\otimes e^6)\;,
\end{equation}
with moduli $R_1$ and $R_2$.
%%%
A basis of $Sp(2)$--invariant two- and three-forms can be found from Eq.~\eqref{Jinv} and is given by
\begin{equation}
e^{12}+e^{56}\; ,\quad e^{34}\; ,\quad e^{135}+e^{146}-e^{236}+e^{245}\; ,\quad e^{136}-e^{145}+e^{235}+e^{246}\; .
\end{equation}
The most general linear combinations of these forms, defining an $Sp(2)$--invariant $SU(3)$ structure are
\begin{align}
  J&=R_1^2\;e^{12}-R_2^2\;e^{34}+R_1^2\;e^{56}\;,\label{JSp2}\\
  \Omega&=R_1^2R_2\left((e^{136}-e^{145}+e^{235}+e^{246})+\mathrm{i} \;(e^{135}+e^{146}-e^{236}+e^{245})\right)\label{OmegaSp2}\;.
\end{align}
As before, we can find a basis of $Sp(2)$--invariant forms which satisfies the defining properties of half-flat mirror manifolds, as outlined in Section~\ref{mirrorhalflat}. It turns out, the correct choice is
\begin{align}\label{eq_sp2basisstart}
  \omega_1 &=\frac{1}{(2\pi)}\left(e^{12}+2e^{34}+ e^{56}\right)
& 
\tilde\omega^1 &=\frac{(2\pi)}{6\,{\cal V}}\left(e^{1234}+2 e^{1256}+ e^{3456}\right)
\\
 \omega_2 &= \frac{1}{3(2\pi)}\left(e^{12}-e^{34}+e^{56}\right)
& 
\tilde\omega^2 &=\frac{(2\pi)}{{\cal V}}\left(e^{1234}-e^{1256}+e^{3456}\right)
\\
  \alpha_0 &=\frac{(2\pi)}{4{\cal V}}\left(e^{136}-e^{145}+e^{235}+e^{246}\right)
& 
\beta^0 &=\frac{1}{(2\pi)}\left(e^{135}+e^{146}-e^{236}+e^{245}\right)\;,\label{eq_sp2basisend}
\end{align}
with the volume ${\cal V}$ of the coset space defined as in Eq.~\eqref{vol}. These forms indeed satisfy the relevant relations~\eqref{hfmdef} for half-flat mirror manifolds provided the torsion parameters are set to
\begin{equation}
  e_1=0\;, \quad e_2=1\;.\label{eSp2}
\end{equation}
The torsion classes are given by~\cite{Chatzistavrakidis:2009mh},
\begin{eqnarray}
W_1^+ &=& \frac{2\left(2R_1^2+R_2^2\right)}{3R_1^2R_2}\;,\\
W_2^+ &=& \frac{4}{3R_1^2R_2}\left[R_1^2\left(R_1^2-R_2^2\right)e^{12}+2R_2^2\left(R_1^2-R_2^2\right)e^{34}+R_1^2\left(R_1^2-R_2^2\right)e^{56}\right]\; .
\end{eqnarray}
When the two radii are equal, $R_1=R_2\equiv R$, they simplify to
\begin{equation}
W_1^+ =\frac{2}{R}\;,\quad W_2^+=0\; ,
\end{equation}
which correspdonds to a nearly-Kahler $SU(3)$ structure, as before. Expanding $J$ in Eq.~\eqref{JSp2} in terms of the forms $\omega_\imath$, we obtain the Kahler moduli fields, 
\begin{equation}
  v^1=\frac{(2\pi)}{3}\left(R_1^2-R_2^2\right)\; ,\quad v^2=(2\pi)\left(2R_1^2+R_2^2\right)\; .
\end{equation}
The form $\omega_1$ is closed while $\omega_2$ is not, so we expect $v^1$ to be massless and $v^2$ to be heavy. From the torsion parameters~\eqref{eSp2} the superpotential~\eqref{W} is given by $W=T^2$ which confirms this expectation.  As before, there are no ``complex structure moduli" for this coset space.

%%%%%%
\subsubsection{$G_2/SU(3)$}
Details of the group theory are explicitly given in appendix \ref{appendix_g2structureconstants}. Following the same procedure as in the previous two cases, the most general $G_2$ invariant metric turns out to be
\begin{equation}
  ds^2=R^2\;(e^1\otimes e^1+e^2\otimes e^2+e^3\otimes e^3+e^4\otimes e^4+e^5\otimes e^5+e^6\otimes e^6),
\end{equation}
where $R$ is the only modulus. The $G_2$ invariant $SU(3)$-structure forms are given by
\begin{align}\label{G2J}
  J&=R^2\;(-e^{12}+e^{34}+e^{56}),\\
  \Omega&=R^3\left(\left(e^{136}+e^{145}-e^{235}+e^{246}\right)+ \mathrm{i}\,\left(e^{135}-e^{146}+e^{236}+e^{245}\right)\right)\; .
\end{align}
The basis 
\begin{align}
  \omega_1&=\frac{1}{2\sqrt{3}(2\pi)}\left(-e^{12}+e^{34}+e^{56}\right)& \tilde\omega^1&=\frac{2(2\pi)}{\sqrt{3}{\cal V}}\left(e^{1234}+e^{1256}-e^{3456}\right)\\
  \alpha_0&=\frac{(2\pi)}{4{\cal V}}\left(e^{136}+e^{145}-e^{235}+e^{246}\right)& \beta^0&=\frac{1}{(2\pi)}\left(e^{135}-e^{146}+e^{236}+e^{245}\right)\;,
\end{align}
satisfies the half-flat mirror condition~\eqref{mirrorhalflat} with the intrinsic torsion parameter given by
\begin{equation}
  e_1=1.
\end{equation}
The only non-vanishing torsion class is~\cite{Chatzistavrakidis:2009mh},
\begin{equation}
W_1^+ = \frac{4}{\sqrt{3}R}\;.
\end{equation}
The single Kahler modulus
\begin{equation}
  v^1=2\sqrt{3}(2\pi)\; R^2,
\end{equation}
is a heavy mode since $\omega^1$ is not closed or, equivalently, since the superpotential is given by $W=T^1$. Once more, there are no complex structure moduli. 

%%%%%%%%%%%%%%%%%%%%%%%%%%%%%%%%%%%%%%%%%%%%%%%%%%%%%%%%%%%%%%%%%%%%%%%%%%%%%%%%%%%%%%%%%%%%%%%%%%%%%%%%%%%%%%%%%%%%%%%%%%%%%%%%%%%%%%%%%%%%%%%%
\section{Vector bundles}\label{section_vectorbundles}
\setall

So far, we have set the scene by presenting the gravitational sector of certain non-Calabi-Yau heterotic compactifications. We now come to the main point of the paper which is the construction of gauge fields associated to these compacifications. To date, gauge fields in heterotic non-Calabi-Yau compactifications have been mainly addressed in a generic way, without providing explicit bundles and connections. Obviously, this restricts phenomenological applications of non-Calabi-Yau models considerably. 
One reason for this is the lack of suitable example manifolds on which to construct gauge bundles. In the case of non-Calabi-Yau manifolds without an integral complex structure, the case considered in this paper, an added complication is that powerful tools from algebraic geometry which are essential in Calabi-Yau model building cannot be directly applied. (An interesting new class of examples where one may be able to circumnavigate this problem has been found in Ref.~\cite{Larfors:2011zz,Larfors:2010wb}.) In the present paper, we focus on a small class of half-flat coset manifolds suitable for heterotic compactifications, which have the advantage of allowing for an explicit computation of most relevant gauge field quantities. Discussion about SU(3)-equivariant pseudo-holomorphic bundles over $SU(3)/U(1)^2$ can also be found in~\cite{Popov:2010rf}.

In this section, the basic mathematical methods for constructing bundles and connections on coset spaces and evaluating their properties will be explained. In particular, we will concentrate on how to construct line bundles on coset spaces. These can be used as building blocks to construct the higher rank bundles which are typically of interest in heterotic compactifications. We will also show how the index of bundles -- giving the number of chiral families in the low-energy theory -- can be computed from the Atiyah-Singer index theorem. In the next section, these general constructions will be applied to our particular coset examples. As we will see, explicit gauge connections and their associated field strengths for the relevant bundles can be written down for these spaces. It is this feature, facilitated by the group structure of the manifolds, which allows us to check all relevant properties required for heterotic vacua.

\subsection{Associated vector bundles and line bundles}\label{assbundles}
We have mentioned before that the group $G$ can also be viewed as a principle bundle $G=G(G/H , H)$ over the coset space $X=G/H$. This observation is the starting point for constructing vector bundles $V$ over $G/H$. It is well-known, that for each representation $\rho$ of $H$ on a vector space $F$, there is a vector bundle $V=V(G/H,F)$ over $G/F$, with typical fiber $F$, which is associated to the principle bundle $G$. More explicitly, this vector bundle can be constructed as follows. We start with the trivial vector bundle $G\times F$ over $G$, where the group $H$ acts on the fiber $F$ via the representation $\rho$. On this vector bundle, we can introduce the equivalence relation
\begin{equation}
 (g,\xi) \sim (g\cdot h,\rho(h^{-1})\xi)\; .
\end{equation}
The vector bundle $V_\rho$ over the coset $G/H$ is then defined as the set of equivalence classes under this relation. Hence, for every representation $\rho$ of $H$ we have a corresponding vector bundle $V_\rho$ over the coset $G/H$ which is associated to the principle bundle $G$. A particularly useful fact for our purposes is that a connection on the principal bundle uniquely induces a connection on every associated vector bundle. This leaves us with finding a connection on $G(G/H,H)$ and in the spirit of chapter \ref{chapter3} we require this connection to be $G$-invariant. Fortunately, it is known \cite{Nomizu} that the $G$-invariant connections of the principal bundle $G(G/H,H)$ are in one-to-one correspondence with reductive decompositions of $G$ and are explicitly given by
\begin{equation}
A=\varepsilon^iH_i \; .
\end{equation} 
We recall, that the $H_i$ are a basis of the Lie-algebra of $H$ and the one-forms $\varepsilon^i$ on the coset have been defined in Eq.~\eqref{Vdef}. The induced connection, $A_{(\rho)}$, on the associated vector bundle $V_\rho$ is then given by
\begin{equation}\label{eq_ginvconncection}
  A_{(\rho)}=\varepsilon^i\rho(H_i)\;.
\end{equation}
The curvature $F=dA+A\wedge A$ of this connection~\footnote{For simplicity of notation we will drop the index $\rho$ from hereon.} can be computed from the Maurer-Cartan structure equations~\eqref{d}. This leads to
\begin{equation}\label{F}
  F=-\frac{1}{2}f_{ab}^{\phantom{ab}i} \rho(H_i) e^a\wedge e^b\;.
\end{equation}
Note this curvature is independent of $\varepsilon^i$ and can be expressed solely in terms of the vielbein forms $e^a$ as a direct consequence of reductiveness, that is, of the structure constants satisfying~\eqref{eq_reductivity}. This fact is of considerable practical importance since it means that all subsequent calculations can be performed ``algebraically", merely based on the knowledge of structure constants. 

We would like to mention two specific types of associated vector bundles which will be relevant for the subsequent discussion. The first is obtained by choosing the representation
\begin{equation}\label{torsionconnection}
  \rho(H_i)_{b}^{\phantom{cb}a}=f_{ib}^{\phantom{ib}a}\;,
\end{equation}
that is, $\rho$ is induced by the adjoint representation of $G$. The corresponding bundle is the tangent bundle of $G/H$ and the gauge field defined by the above choice of representation provides a connection with torsion on this bundle. 

The second type arises for one-dimensional representations $\rho$ of $H$. Applying the above formalism to such representations leads to line bundles and connections on them. One choice which is always possible is of course the trivial representation of $H$. However, in this case the associated line bundle is simply the trivial line bundle ${\cal O}_X$. Fortunately, for two of our examples, the corresponding sub-groups $H$ allow for non-trivial one-dimensional representations so that we can generate more interesting line bundles, $L$. Since we know the curvature form of these line bundles it is possible to explicitly work out their first Chern class
\begin{equation}
 c_1(L)=\frac{i}{2\pi}[F]=p^r \omega_r\; .
\end{equation} 
Here the square bracket denotes the cohomology class in $H^2(X)$. The last part of the equation is a linear combination of a suitable basis, $\{\omega_r\}$, of $H^2(X)$, to be determined explicitly for our examples, with integer coefficients $p^r$. The line bundle $L$ is uniquely characterised by its first Chern class or, equivalently, by the integer vector ${\bf p}=(p^r)$, and will also be denoted as $L={\cal O}_X({\bf p})$. These line bundles will be used as building blocks for higher-rank bundles. In particular, we will consider sums of $n$ line bundles
\begin{equation}
  V=\bigoplus_{i=1}^nL_i\quad {\rm where} \quad L_i={\cal O}_X({\bf p}_i)\;. \label{V}
\end{equation}
For such line bundle sums we require a vanishing total first Chern class, $c_1(V)=0$, which means the integers $p_i^r$ must satisfy
\begin{equation}
 \sum_{i=1}^np_i^r=0
\end{equation}
for all $r$. This guarantees that the structure group of $V$ is contained in $S(U(1)^n)$. For $1<n\leq 8$ this allows for an embedding into one of the $E_8$ factor of the gauge group via the sub-group chain $S(U(1)^n)\subset SU(n)\subset E_8$. The low-energy gauge group in this $E_8$ sector is the commutant of the bundle structure group within $E_8$, as usual. For $S(U(1)^n)$ with $n=3,4,5$ this commutant is given by $S(U(1)^3)\times E_6$, $S(U(1)^4)\times SO(10)$ and $S(U(1)^5)\times SU(5)$, respectively, and, therefore, contains phenomenologically interesting GUT groups as its non-Abelian part. 

For consistent heterotic vacua the gauge bundle needs to satisfy further requirements. First of all, we need to satisfy the supersymmetry conditions~\eqref{HYM}. Since we know the gauge field strengths $F$ on our bundles, as well as the $SU(3)$ structure forms $(J,\Omega)$ these conditions can be checked explicitly and this is what we will do for our examples. It will turn out that both the connection~\eqref{torsionconnection} as well as line bundle sums can satisfy the supersymmetry conditions.

In addition, we need to satisfy the integrability condition~\eqref{biancoho} for the Bianchi identity and we now turn to a discussion of this task.
%%%%
\subsection{Bianchi identity}
We recall from Eq.~\eqref{biancoho} that the integrability condition for the Bianchi identity reads
\begin{align}\nonumber
  \left[{\rm tr}R\wedge R\right]= \left[{\rm tr}F\wedge F\right]+\left[{\rm tr}\tilde F\wedge \tilde F\right]\;,\label{biancoho1}
\end{align}
where the square bracket indicates cohomology classes in $H^4(X)$. Here, $R$ is the curvature tensor of the coset space $X$ and $F$, $\tilde{F}$ are the field strengths in the two $E_8$ sectors, corresponding to observable and hidden bundles $V$ and $\tilde{V}$. In terms of characteristic classes this can be written as
\begin{equation}
p_1(TX)=2\left({\rm ch}_2(V)+{\rm ch}_2(\tilde{V})\right)\; . \label{bc}
\end{equation}
In practice, we will write those classes as a linear combination of a basis, $\{\tilde{\omega}^r\}$ of $H^4(X)$, dual to our earlier bases, $\{\omega_r\}$ of the second cohomology. The relation between those two basis sets can be written as
\begin{equation}
 \omega_r\wedge\omega_s=d_{rst}\tilde{\omega}^t\; , \label{isec}
\end{equation}
where the $d_{rst}$  are the analogous to triple intersection numbers. The numbers $d_{rst}$ will be explicitly determined for our examples.

Both, for the computation of the intersection numbers and the left-hand side of the anomaly condition~\eqref{bc}, we require the curvature tensor of the the coset space.  The Levi-Civita connection one-form, $\omega^a_{\;\;b}$, associated to the vielbein $e^a$ on the coset space is determined by the standard relations $de^a+\omega^a_{\phantom{a}b}\wedge e^b=0$ and $\omega_{ab}=-\omega_{ba}$. For reductive homogeneous spaces it is given by~\cite{MuellerHoissen:1987cq}
\begin{equation}
  \omega_{cb}^{\phantom{cb}a}e^c=D_{cb}^{\phantom{cb}a}e^c+f_{ib}^{\phantom{ib}a}\varepsilon^i\;\;\mbox{where }\;\;
  D_{cb}^{\phantom{cb}a}=\frac{1}{2}f_{cb}^{\phantom{cb}a}-\frac{1}{2}(g^{am}f_{cm}^{\phantom{cm}n}g_{nb}+g^{am}f_{bm}^{\phantom{bm}n}g_{cn})\; ,\label{omega}
\end{equation}
which leads to the curvature two-form $R^a_{\phantom{a}b}=\frac{1}{2}R^a_{\phantom{a}bcd}\,e^c\wedge e^d$ with
\begin{equation}
 R^a_{\phantom{a}bcd}=-f_{cd}^{\phantom{cd}i}f_{ib}^{\phantom{ib}a}-f_{cd}^{\phantom{cd}m}D_{mb}^{\phantom{mb}a}+D_{cm}^{\phantom{cm}a}D_{db}^{\phantom{db}m}-D_{dm}^{\phantom{dm}a}D_{cb}^{\phantom{cb}m}\;. \label{R}
\end{equation}
From this result, we can work out the first Pontryagin class for our examples and write it as $p_1(TX)=p_{1r}(TX)\tilde{\omega}^r$ in terms of our basis for the fourth cohomology. The crucial input for the computation of the intersection numbers is the volume~\eqref{vol} of the coset space. This can be obtained by using the generalized Gauss-Bonnet theorem\cite{MuellerHoissen:1987cq}
\begin{equation}
  \chi(X)=\int_X\gamma(TX)\;,
\end{equation}
where $\chi$ is the Euler characteristic and $\gamma$ the Euler form of $X$. The latter can be expressed in terms of the Riemann curvature tensor and in six dimensions it is given by
\begin{equation}
  \gamma(TM)=\frac{-1}{2^3(2\pi)^33!}\sum\epsilon^{a_1...a_6}R_{a_1a_2}\wedge...\wedge R_{a_5a_6}\equiv \frac{1}{v} e^1\wedge...\wedge e^6\;.
\end{equation}
Here the last equality defines the quantity $vi$. In practice, we work out $v$ by inserting the above result for the curvature tensor and by re-writing the resulting expression as a constant times $e^1\wedge\ldots\wedge e^6$. The volume is then given by
\begin{equation}\label{vchi}
  {\cal V}= v\chi\;.
\end{equation}

Which choice of gauge bundle should we make in order to satisfy the anomaly condition~\eqref{bc} for a given manifold, that is, a given first Pontryagin class on the left-hand side? One obvious attempt would be to set the ``observable" gauge field $F$ equal to the above curvature, while choosing the hidden curvature to be trivial. This would obviously satisfy the Bianchi identity~\eqref{bianchi}, not just in cohomology, but point-wise on the coset space for a vanishing three-form $H$. This choice is the analogue of the ``standard embedding" traditionally used in heterotic Calabi-Yau compactifications. In the present context, the problem with this choice is that, for our coset spaces, the curvature~\eqref{R} does not satisfy the supersymmetry conditions~\eqref{HYM} required for the gauge fields. Hence, we cannot choose a standard embedding in the conventional sense.

However, a related choice, somewhat reminiscent of the standard embedding, is possible. We can choose the observable gauge field specified by \eqref{torsionconnection} on the tangent bundle while the hidden gauge field is trivial. This will satisfy the anomaly condition~\eqref{bc} since both, \eqref{omega} and \eqref{torsionconnection}, provide connections on the same bundle and will, hence, result in the same topological characteristics. Also, as we have mentioned earlier, the gauge field connection defined by \eqref{torsionconnection} can indeed satisfy the supersymmetry conditions~\eqref{HYM} for our coset spaces, as we will show. Hence, this choice leads to a consistent and supersymmetric vacuum. However, since the curvature forms~\eqref{F}, \eqref{torsionconnection} and \eqref{R} are not the same (in fact, the former is equal to the first term in the latter) the right-hand side of the Bianchi identity does not vanish point-wise and a non-zero $H$-field will be required at order $\alpha'$. For this reason it might not be appropriate to refer to this choice as ``standard embedding".

However, we would like to work with more general gauge fields, rather than special choices resembling the standard embedding. Our focus will be on the simplest such class with Abelian structure groups. This means that the associated vector bundles are sums of line bundles as in Eq.~\eqref{V}. More precisely, we will allow for both an observable bundle $V$ and a hidden bundle $\tilde{V}$ of this kind, that is,
\begin{equation}
 V=\bigoplus_{i=1}^n{\cal O}_X({\bf p}_i)\;, \quad \tilde{V}=\bigoplus_{j=1}^m{\cal O}_X(\tilde{\bf p}_i)\; .
 \end{equation}
We demand vanishing first Chern classes, $c_1(V)=c_1(\tilde{V})=0$, to allow for an embedding into the two $E_8$ factors. This translates into
\begin{equation}
 \sum_{i=1}^np_i^r=\sum_{j=1}^m\tilde{p}_j^r=0
\end{equation}
for all $r$. Using additivity of the Chern character and the fact that ${\rm ch}_2(L)=c_1(L)^2/2$ for a line bundle $L$, together with Eq.~\eqref{isec}, we find for the second Chern character ${\rm ch}_{2}(V)={\rm ch}_{2r}(V)\tilde{\omega}^r$ that
\begin{equation}
{\rm ch}_{2r}(V)=\frac{1}{2}d_{rst}\sum_{i=1}^np_i^sp_i^t\; , \label{ch2V}
\end{equation}
and analogously for $\tilde{V}$. With this result, the anomaly condition~\eqref{bc} can be written as
\begin{equation}
d_{rst}\left(\sum_{i=1}^np_i^sp_i^t+\sum_{j=1}^m\tilde{p}_j^s\tilde{p}_j^t\right)=p_{1r}(TX)\; .
\end{equation} 
 
\subsection{Index Formula}
One of the most basic topological invariants of bundles is the index which gives the chiral asymmetry of zero modes of the Dirac operator and, hence, the net number of families in the four-dimensional theory. The index can be computed from the Atiyah-Singer index theorem \cite{Nash:1991pb} which involves the A-roof genus
\begin{equation}
 \hat{A}(X)=1-\frac{1}{24}p_1(TX)+\dots\; .
\end{equation}
of the manifold $X$. For a bundle $U$ on a six-dimensional manifold $X$ the index theorem then takes the form
\begin{equation}
 {\rm ind}(U)=-\int_X\hat A(X)\wedge {\rm ch}(U)=-\int_X\left[{\rm ch}_3(U)-\frac{1}{24}p_1(TX){\rm ch}_1(U)\right]\; . \label{AS}
\end{equation}
For a line bundle, $L$, we have ${\rm ch}_3(L)=c_1(L)^3/6$, where $c_1(L)=c_1^r(L)\omega_r$ is the first Chern class of $L$. Inserting this, together with the definition~\eqref{isec} of the intersection numbers, into the index formula~\eqref{AS} leads to
\begin{equation}
{\rm ind}(L)=-\frac{1}{6}d_{rst}c_1^r(L)c_1^s(L)c_1^t(L)+\frac{1}{24}p_{1r}(TX)c_1^r(L)\; .
\end{equation}
In this paper, we mainly consider sums of line bundles $V=\bigoplus_{i=1}^nL_i$, where $L_i={\cal O}_X({\bf p}_i)$, with vanishing first Chern class, $c_1(V)=0$. For such bundles the above formula simplifies to
\begin{equation}
 {\rm ind}(V)=-\frac{1}{6}d_{rst}\sum_{i=1}^np_i^rp_i^sp_i^t\; . \label{indV}
\end{equation} 
Hence, we only need to know the intersection numbers $d_{rst}$ of the manifold $X$, together with the integers, $p_i^r$ characterizing the line bundles in order to work out the index. We will also consider some non-Abelian bundles, $V$ with vanishing first Chern class. In this case, it is convenient to express the index~\eqref{AS} in terms of the curvature, $F$, of $V$. This leads to
\begin{equation}\label{index}
  {\rm ind}(V)=\frac{i}{6(2\pi)^3}\int_X{\rm tr}\left( F\wedge F\wedge F\right)\; .
\end{equation}  

%%%%%%
\section{Bundles on coset spaces}\setall
We would now like to apply the above bundle constructions to the three coset spaces introduced earlier. Wherever possible, our focus will be on line bundle sums, although we will discuss some specific non-Abelian bundles as well.

\subsection{$SU(3)/U(1)^2$}
Let us first specify some of the required coset properties for this case. The $SU(3)$ generators $\{T_A\}=\{K_a,H_i\}$ are split into the six coset generators, $K_a$,  $a=1,\ldots ,6$, given by the non-diagonal Gell-Mann matrices and the two generators $H_i$,  $i=7,8$ of the sub-group $U(1)^2$, given by the two diagonal Gell-Mann matrices. The explicit matrices and the associated structure constants are presented in Appendix~\ref{appendix_su3structureconstants}. The second Betti number of this coset space is two, so we have two basis forms $\{\omega_r\}$ and $\{\tilde{\omega}^r\}$, where $r=1,2$, each for the second and fourth cohomology, respectively. They are explicitly given by the forms in Eqs.~\eqref{eq_su3basisstart}, \eqref{eq_su3end} introduced earlier. From Eq.~\eqref{vchi} one can work out the volume~\eqref{vol} of the coset which is given by
\begin{equation}
  {\cal V}=4(2\pi)^3\;.
\end{equation}
This results in the following intersection numbers.
\begin{equation}
 d_{111}=6\; ,\quad d_{112}=3\; ,\quad d_{122}=1\; ,\quad d_{222}=0\; . \label{su3isec}
 \end{equation}
For the first Pontryagin class of the tangent bundle we find
\begin{equation}
 p_1(TX)=0\; .
\end{equation}  

Let us first discuss possible non-Abelian bundles. Using the explicit structure constants from Appendix~\ref{appendix_su3structureconstants}  we can verify that the Levi-Civita curvature~\eqref{R} does not satisfy the supersymmetry equations~\eqref{HYM} and, hence, cannot be used as a gauge curvature. Let us consider the supersymmetry conditions for associated bundles, specified by representations, $\rho$, of the sub-group $H$ as introduced in Section~\ref{assbundles}. First, it can be checked that the  constraint $\Omega\,\neg\, F=0$ is always trivially satisfied. The constraint $J\,\neg\, F=0$ implies explicitly that
\begin{equation}
J^{ab}f_{ab}^{\phantom{ab}i}\rho(H_i)=\left(\frac{2}{R_1^2}-\frac{1}{R_2^2}-\frac{1}{R_3^2}\right)\rho(H_7)+\left(\frac{\sqrt{3}}{R_3^2}-\frac{\sqrt{3}}{R_2^2}\right)\rho(H_8)=0\;.
\end{equation}
In general, the two representation matrices are linearly independent, so we have two constraints on the moduli which are solved by
\begin{equation}\label{equalR}
  R_1^2=R_2^2=R_3^2\equiv R^2\;.
\end{equation}
Hence, all associated bundles are supersymmetric on the nearly-Kahler locus of the moduli space. In particular, this applies to the connection~\eqref{torsionconnection}. However, from Eq.~\eqref{index}, its index vanishes as one would expect for an associated vector bundle which corresponds to a real representation of the group $H$. Hence, it is not of particular interest from a physics point of view. 

Associated bundles which correspond to irreducible representations of the sub-group $H$ can be viewed as ``building blocks" for general associated bundles. In the present case, the sub-group $H=U(1)^2$ is Abelian so that all irreducible representations are one-dimensional and, hence, lead to line bundles. We characterize an irreducible representation $\rho$ by a pair, $(p,q)$ of integer charges and, more specifically, define the representation by 
\begin{equation} 
  \rho(H_7)= -i(p+q/2)\; ,\quad\rho(H_8)= -iq/(2\sqrt{3}),
\end{equation}
From Eq.~\eqref{F} this means the associated curvature form is given by
\begin{equation}
 \frac{F}{2\pi}=-ip\omega_1-iq\omega_2\; ,
\end{equation}
and the first Chern class of the associated line bundle, $L$ is $c_1(L)=p\omega_1+q\omega_2$. From our earlier discussion this means $L$ should be identified with ${\cal O}_X(p,q)$. Taking the observable and hidden bundles, $V$ and $\tilde{V}$,  as sums of line bundles with vanishing first Chern class we can, therefore write
\begin{equation}
 V=\bigoplus_{i}{\cal O}_X(p_i,q_i)\; ,\quad\tilde{V}=\bigoplus_{j}{\cal O}_X(\tilde{p}_j,\tilde{q}_j)\; , \label{Vsu3}
 \end{equation}
where
\begin{equation}
 \sum_{i}^np_i=\sum_{j}^nq_j=0\; , \label{su3c10}
\end{equation} 
and similarly for $\tilde{p}_i$ and $\tilde{q}_i$. Hence, each such sum of $n$ line bundles is determined by the $2n$ integers $p_i$, $q_i$, subject to the constraints~\eqref{su3c10}. For the second Chern character, relative to the basis $\{\tilde{\omega}^1,\tilde{\omega}^2\}$,  we find from Eqs.~\eqref{ch2V} and \eqref{su3isec}
\begin{equation}
 {\rm ch}_2(V)=\left(\sum_{i}(3p_i^2+\frac{1}{2}q_i^2+3p_iq_i),\sum_{i}(p_iq_i+\frac{3}{2}p_i^2)\right)\; ,
 \end{equation}
and similarly for $\tilde{V}$. Analogously, Eqs.~\eqref{indV} and \eqref{su3isec} lead to the expression
\begin{equation}
{\rm ind}(V)=-\sum_i\left(p_i^3+\frac{1}{2}p_iq_i(q_i+3p_i)\right)\; . \label{indVsu3}
\end{equation}
for the index of $V$. Again, the supersymmetry equations~\eqref{HYM} can be solved by the constraint,
\begin{equation}
  R_1^2=R_2^2=R_3^2\equiv R^2\;.
\end{equation}
From the above result for the second Chern character (remembering that the first Pontryagin class for this coset vanishes) the two components of the anomaly cancelation condition~\eqref{bc} can be written as
\begin{eqnarray}\label{secondchernzero1}
  \sum_i(3p_i^2+\frac{1}{2}q_i^2+3p_iq_i)+\sum_j(3\tilde p_j^2+\frac{1}{2}\tilde q_j^2+3\tilde p_j\tilde q_j)=0\;, \\ \label{secondchernzero2}
  \quad \sum_i(p_iq_i+\frac{3}{2}p_i^2)+\sum_j(\tilde p_j\tilde q_j+\frac{3}{2}\tilde p_j^2)=0\;.
\end{eqnarray}
We have now collected all results required for basic model building on this coset. The problem is to choose observable bundles, $V$, with ${\rm rk}(V)=3,4,5$ specified by integers $p_i$, $q_i$ and corresponding hidden bundles, $\tilde{V}$, with ${\rm rk}(\tilde{V})=2,\ldots ,8$ specified by integers $\tilde{p}_j$, $\tilde{q}_j$ subject to the following constraints.
\begin{itemize}
\item The first Chern classes of $V$ and $\tilde{V}$ vanish, that is, Eqs.~\eqref{su3c10} are satisfied.
\item The anomaly conditions~\eqref{secondchernzero1} and \eqref{secondchernzero2} are satisfied.
\item The index~\eqref{indVsu3} of the observable bundle $V$ equals three to obtain a GUT model with three net families.
\end{itemize}
It is clear that there are many possible solutions to these constraints and a systematic study of all model building options will be presented in a forthcoming publication~\cite{inprep}. Here, we merely present a number of examples given in Table~\ref{tab1}.
\begin{table}[h]
\begin{center}
\begin{tabular}{|l|l|l|l||l|l|} 
\hline
rank $n$&$p_i$ & $q_i$ & $\#\;{\rm generations}$ & $\tilde p_i$ & $\tilde q_i$\\ \hline \hline
3&(-1,-1,2)& (0,3,-3)&3&(0)&(0)\\ \hline
3&(-2,0,2)& (1,1,-2)&3&(2,1,-1,-1,-1)&(-4,-3,3,2,2)\\ \hline
3&(-2,-1,3)& (1,2,-3)&3&(2,1,0,-1,-2)&(-4,-3,-1,4,4)\\ \hline
4&(-2,-1,1,2)& (1,2,-1,-2)&3&(1,1,1,-1,-2)&(-1,-3,-3,3,4)\\ \hline
4&(-2,0,1,1)& (1,2,-2,-1)&3&(1,1,1,-1,-2)&(-2,-2,-2,2,4)\\ \hline
4&(-1,0,0,1)& (-1,1,1,-1)&3&(2,1,1,-2,-2)&(-3,-1,-3,4,3)\\ \hline
5&(-1,0,0,0,1)& (-1,1,1,1,-2)&3&(-3,-1,1,1,2)&(4,3,-2,-1,-4)\\ \hline
5&(-2,0,0,0,2)& (1,-2,1,2,-2)&3&(2,2,0,-2,-2)&(-3,-4,-1,4,4)\\ \hline
5&(-1,-1,-1,1,2)& (-1,2,2,-1,-2)&3&(1,1,1,-1,-2)&(-1,-2,-3,2,4)\\ \hline
%3,4,5&many more&many more&3&\dots& \dots\\ \hline
\end{tabular}
\parbox{6in}{\caption{\it\small Sample of three generations models with base space $SU(3)/U(1)^2$. Observable and hidden bundles are specified by the integers $(p_i,q_i)$ and $(\tilde p_i, \tilde q_i)$, respectively, as in Eq.~\eqref{Vsu3}. The rank of the hidden bundle has been taken to be five in all the cases but solutions with different ranks exist.}}\label{tab1}
\end{center}
\end{table}

We have seen earlier that the torsion connection~\eqref{torsionconnection} on the tangent bundle, while supersymmetric, has a vanishing index since it is associated to a real representation. It was, therefore, not suitable as a ``standard embedding". A related, complex representation can be defined by considering $H$ as a sub-group of $SU(3)$ and by choosing the representation $\rho$ which is induced by the fundamental representation of $SU(3)$. This means setting 
\begin{equation}
 \rho(H_7)=\sqrt{3}\lambda_8\; ,\quad \rho(H_8)=\sqrt{3}\lambda_3\; . \label{sesu3}
\end{equation}
The associated bundle for this representation has rank three and is, in fact, a sum of three line bundles. It turns out that it corresponds to the  example in the first row of Table~\ref{tab1}. For this choice, the anomaly condition is satisfied for a trivial hidden bundle and the chiral asymmetry, three in this case, equals half the Euler number of this manifold. Hence, this bundle has two of the main characteristics of the standard embedding. Note, however, that it does not lead to a vanishing right-hand side of the Bianchi identity~\eqref{bianchi} and, therefore, the model receives corrections at order $\alpha'$. 

\subsection{$Sp(2)/SU(2)\times U(1)$}
The generators $\{T_A\}=\{K_a,H_i\}$ of $Sp(2)$ consist of six coset generators, $K_a$, $a=1,\ldots ,6$ and four generators $H_i$, $i=7,\ldots , 10$ of the sub-group $SU(2)\times U(1)$. The explicit matrices and associated structure constants are listed in Appendix~\ref{appendix_sp2structureconstants}. The second Betti number of this coset space is one and the second and fourth cohomology are spanned by the forms $\omega_1$ and $\tilde{\omega}^1$ given in Eq.~\eqref{eq_sp2basisstart}. For the volume~\eqref{vol} one finds from Eq.~\eqref{vchi}
\begin{equation}
  {\cal V}=\frac{(2\pi)^3}{12}\; .
\end{equation}
The single intersection number and the first Pontryagin class are given by
\begin{equation}
 d_{111}=1\; ,\quad p_1(TX)=4\tilde{\omega}^1\; .
\end{equation}
As for the $SU(3)$ case one can verify from the structure constants in Appendix~\eqref{appendix_sp2structureconstants} that the Levi-Civita curvature does not satisfy the supersymmetry conditions. For associated bundles with representation $\rho$ the constraint $\Omega\,\neg\, F=0$ is trivialy satisfied while the constraint $J\,\neg\, F=0$ implies 
\begin{equation}
J^{ab}f_{ab}^{\phantom{ab}i}\rho(H_i)=\left(\frac{4}{R_1^2}-\frac{4}{R_2^2}\right)\rho(H_{10})=0\;.
\end{equation}
This is solved in the region of moduli space where
\begin{equation}
  R_1^2=R_2^2\equiv R^2\;.
\end{equation}
Hence, all associated bundles satisfy the supersymmetry conditions in the nearly-Kahler part of the moduli space. In particular, this applies to the connection~\eqref{torsionconnection}. However, as before, it has a vanishing index and is, therefore, of limited interest. 

Line bundles $L={\cal O}_X(p)$ are characterized by a single integer $p$. They can be constructed as associated bundles by choosing representations $\rho$ of the sub-group $SU(2)\times U(1)$ which are trivial on the $SU(2)$ part and have $U(1)$ charge $p$. Explicitly, this means
\begin{equation}
  \rho(H_7)=0\; ,\quad\rho(H_8)= 0\; ,\quad\rho(H_9)=0\; ,\quad\rho(H_{10})=ip\; .
\end{equation}
The associated field strength is $F/(2\pi)=-ip\omega_1$ which shows that the associated bundle has first Chern class $c_1(L)=p\omega_1$ and should indeed be identified with ${\cal O}_X(p)$. As before, the observable and hidden bundles $V$ and $\tilde{V}$ are taken as line bundle sums with vanishing first Chern class, so that
\begin{equation}
 V=\sum_i{\cal O}_X(p_i)\; ,\quad \tilde{V}=\sum_j{\cal O}_X(\tilde{p}_j)\; ,\quad\sum_ip_i=\sum_j\tilde{p}_j=0\; . \label{Vsp2}
\end{equation} 
For the second Chern character and the index we find
\begin{equation}
 {\rm ch}_2(V)=\frac{1}{2}\sum_ip_i^2\tilde{\omega}^1\; ,\quad {\rm ind}(V)=-\frac{1}{6}\sum_ip_i^3\; , \label{indsp2}
\end{equation}
and similarly for $\tilde{V}$. The anomaly condition now reads
\begin{equation}
  \sum_ip_i^2+\sum_j\tilde p_j^2=4\;.  \label{anomsp2}
\end{equation}
We should now study the model building options in analogy to what we did for $SU(3)/U(1)^2$. We need to choose bundles $V$ and $\tilde{V}$, specified by integers $p_i$ and $\tilde{p}_j$ as in \eqref{Vsp2} which satisfy the anomaly condition~\eqref{anomsp2} and lead to an index~\eqref{indsp2} of three so that we obtain three chiral GUT families. However, unlike for the previous case, the combination of these conditions is quite restrictive. A search over all integers $p_i$ for ${\rm rk}(V)=3,4,5$ and all integers $\tilde{p}_j$ shows there is only one solution, given by the rank four observable bundle
\begin{equation}
 (p_i)=(1,1,-1,-1)\; ,
\end{equation} 
and a trivial hidden bundle, which satisfies the anomaly condition. Unfortunately, however, this model has vanishing index so is not of physical interest. 

Finally, let us work out the quasi standard-embedding~\eqref{sesu3} for the present case. We recall that this is done by choosing the representation $\rho$ which is induced by the fundamental of $SU(3)$ via the embedding $SU(2)\times U(1)\subset SU(3)$. This means explicitly
\begin{equation}
  \rho(H_7)= -2\lambda_1\; ,\quad\rho(H_8)=-2\lambda_2\; ,\quad\rho(H_9)=-2\lambda_3\; ,\quad\rho(H_{10})=-2\sqrt{3}\lambda_8\; ,
\end{equation}
where $\lambda_i$ are the Gell-Mann matrices as given in Appendix~~\ref{appendix_su3structureconstants}. This choice satisfies the anomaly condition for a trivial hidden bundle. From Eq.~\eqref{index}, we can calculate the index explicitly and we find two chiral families which equals half the Euler number, as expected.

\subsection{$G_2/SU(3)$}
The generators $\{T_A\}=\{K_a,H_i\}$ of $G_2$ consist of the six coset generators $K_a$, $a=1,\ldots ,6$ and the eight generators $H_i$, $i=7,\ldots ,14$ of the sub-group $SU(3)$. The explicit matrices and structure constants are given in Appendix~\ref{appendix_g2structureconstants}. The second Betti number of this coset vanishes so, unfortunately, there are no non-trivial line bundles. Hence, we have to consider non-Abelian gauge fields in this case.

The Levi-Civita connection and the torsion connection~\eqref{torsionconnection} on the tangent bundle have the same properties as for the two previous cases. The former does not satisfy the supersymmetry conditions while the latter does but has a vanishing index.

The quasi standard embedding~\eqref{sesu3} is here obtained by choosing $\rho$ to be the fundamental representation of the $SU(3)$ sub-group. In practice, this means setting
\begin{equation}
\rho(H_i)=- 2\lambda_{i-6}\; .
\end{equation}
Note here that, by our convention, the index $i$ numbering the sub-group generators $H_i$ runs in the range $7,\ldots ,14$.
The anomaly condition is satisfied with a trivial hidden bundle  and the number of generations is half the Euler number and, hence, equal to one.

%%%%%%%%%%%%%%%%%%%%%%%%%%%%%%%%%%%%%%%%%%%%%%%%%%%%%%%%%%%%%%%%%%%%%%%%%%%%%%%%%%%%%%%%%%%%%%%%%%%%%%%%%%%%%%%%%%%%%%%%%%%%%%%%%%%%%%%%%%%%%%%%
\section{Conclusion and outlook}
\setall
In this paper, we have studied the compactification of the heterotic string on six-dimensional coset spaces $G/H$ with particular focus on the three coset spaces $SU(3)/U(1)^2$, $Sp(2)/SU(2)\times U(1)$ and $G_2/SU(3)$. These coset spaces are half-flat and they solve the gravitational sector of the theory in the context of heterotic domain wall vacua. We have shown that the three coset spaces have the structure of half-flat mirror manifolds and, hence, the general results for the gravitational part of the four-dimensional effective theory obtained in Refs.~\cite{Gurrieri:2004dt,Gurrieri:2007jg} can be directly applied. The main purpose of this paper has been to gain a better understanding of the gauge field sector in heterotic half-flat compactifications. 

The group origin of the coset spaces facilitates the construction of gauge bundles and the computation of explicit connections on them. The supergravity equations can, therefore, be checked directly. Specifically, for each representation of the sub-group $H$ one has a vector bundle associated to the principal bundle $G=G(G/H,H)$. For the case $SU(3)/U(1)^2$ the irreducible representations of the sub-group $H=U(1)^2$ leads to line bundles and, in fact, all line bundles on this coset space can be obtained in this way. Since the second Betti number, $b^2$, of this space is two these line bundles are characterised by two integers which correspond to the two charges of $U(1)^2$. The situation is analogous for $Sp(2)/SU(2)\times U(1)$. Taking the $SU(2)$ representation to be trivial the $U(1)$ representations, specified by a single charge, lead to a one-integer family of line bundles, in accordance with $b^2=1$ for this space. The second Betti number of $G_2/SU(3)$ vanishes so there are no non-trivial line bundles on this space as, indeed, there are no non-trivial one-dimensional representations of $H=SU(3)$. Of course, one can also consider higher-dimensional representations and we have presented some examples. One possible choice is the ``fundamental" representation of $H$, that is the representation induced by the fundamental of $SU(3)\supset H$. It turns out that this choice, for all three coset spaces, leads to a quasi standard embedding where the anomaly condition is satisfied for a trivial hidden bundle and the chiral asymmetry is given be half the Euler number of the manifold. 

For the first two coset spaces, we have also shown that consistent vacua can be obtain by suitable sums of line bundles in the observable and hidden sector. The $Sp(2)/SU(2)\times U(1)$ case where line bundles are labeled by only one integer is quite restrictive in this regard and we have been able to find only one consistent model, unfortunately with a vanishing chiral asymmetry. The $SU(3)/U(1)^2$ case, however, allows for many consistent solutions with line bundle solutions and we have presented a number of explicit examples with chiral asymmetry three. 

Specifically, the $SU(3)/U(1)^2$ case requires a more systematic study of all possible line bundle models and work in this direction is underway~\cite{inprep}. More general, non-Abelian bundle constructions, for example based on quotients or extensions of line bundle sums can also be studied in this case and possibly for $Sp(2)/SU(2)\times U(1)$. Recently, a new class of $SU(3)$ structure manifolds which might be suitable for heterotic compactifications has been found by using methods in toric geometry~\cite{Larfors:2011zz,Larfors:2010wb}. It might be interesting to study bundles on this new class of manifolds.

%%%%%%%%%%%%%%%%%%%%%%%%%%%%%%%%%%%%%%%%%%%%%%%%%%%%%%%%%%%%%%%%%%%%%%%%%%%%%%%%%%%%%%%%%%%%%%%%%%%%%%%%%%%%%%%%%%%%%%%%%%%%%%%%%%%%%%%%%%%%%%%%
\section*{Acknowledgements}

We are grateful to Andrei Constantin for early participation in the project. C.\ M.\ would like to thank Maxime Gabella, James Gray and Hwasung Lee for useful discussions and Damien Matti for computing help. C.~M.~ is supported by a Berrow Foundation scholarship in association with Lincoln College Oxford as well as a grant from the Swiss National Science Foundation.
M.~K. would like to thank Christian Paleani for useful discussions. M.~K.\ is supported by the Lamb \& Flag scholarship of St John's College Oxford and by an STFC scholarship. A.~L.~is supported by the EC 6th Framework Programme MRTN-CT-2004-503369 and by the EPSRC network grant EP/l02784X/1.

%%%%%%%%%%%%%%%%%%%%%%%%%%%%%%%%%%%%%%%%%%%%%%%%%%%%%%%%%%%%%%%%%%%%%%%%%%%%%%%%%%%%%%%%%%%%%%%%%%%%%%%%%%%%%%%%%%%%%%%%%%%%%%%%%%%%%%%%%%%%%%%%
\appendix
\section*{Appendix}

%%%%%%%%%%%%
\section{Conventions}\label{Conventions}
\setall
In this appendix we summarize the conventions used throughout the paper. First we present our index conventions, both for 10-dimensional space-time and Lie group generators. Subsequently, we briefly review some standard facts about $SU(3)$ structures which we rely on in the main part of the paper.

%%%%%%
\subsection{Indices}
The ten-dimensional background geometries used in this paper decompose in the following manner,
\begin{equation}
M_{10}=M_3 \times M_1 \times X\;,
\end{equation}
where only $X$ is compact. This is alternatively seen either as a $3+7$ decomposition (where the 7--dimensional space corresponds to $M_1 \times X$) or a $4+6$ decomposition  (where the 4--dimensional space is $M_3 \times M_1$). For this reason we introduce the following sets of indices for the various parts of this decomposition.
\begin{eqnarray}\nonumber 
  10d: && M,N,...=0,1,...,9\\\nonumber
   7d: && m,n,p,...=3,4,...,9\\\nonumber
   6d: && u,v,...=4,5,...,9\\
   4d: && \mu,\nu,...=0,1,2,3\\\nonumber
   3d: && \alpha,\beta,\gamma,...=0,1,2\\\nonumber
  1d: && M=\mu=m=3
\;.
\end{eqnarray}

%%%
Furthermore, in the geometry of coset spaces, we use the following conventions to label the various types of generators.
\begin{eqnarray}\nonumber 
  G: && A,...=1,2,...,{\rm dim}(G)\\
   G/H: && a,b,...=1,2,...,6\\\nonumber
   H: && i,j,...=7,8,...,{\rm dim}(G)
\;.
\end{eqnarray}
Evidently, the indices $a,b,...$ label the internal six-dimensional geometry and therefore are the co-frame (vielbein) indices associated to the above coordinates $u,v,...$.

%%%%%%
\subsection{$SU(3)$ structure}\label{appendix_su3structure}
Generically, the structure group of the frame bundle of a six-dimensional manifold is given by $Gl(6,\mathbb{R})$ but may also be a genuine sub-group of $Gl(6,\mathbb{R})$. This happens if there exist globally defined tensors on the manifold which have to be invariant under the structure group transformation. For the case of an $SU(3)$ structure on a six-dimensional manifold $X$ these tensors are given by a real two-form $J$ and a complex three-form $\Omega$, which are subject to the following algebraic compatibility relations\footnote{Note that the minus sign in the first equation is different from the usual convention in the mathematical literature. It originates from swapping the real and imaginary parts of $\Omega$. This choice has been made in order to be consistent with the conventions in Ref.~\cite{us}.}
\begin{equation}\label{comp}
	J\wedge J\wedge J=-\frac{3}{4}\mathrm{i}\;\Omega\wedge\bar\Omega\;, \quad \Omega\wedge J=0\; .
\end{equation}
It is understood that both sides of the first equation are non-vanishing everywhere and, hence, define a volume form on $X$. Moreover, $J$ and $\Omega$ define a metric on $X$ which we denote by $g$.

%%%
If both forms are closed, the $SU(3)$ structure is integrable and $(X,g)$ has $SU(3)$ holonomy with respect to the Levi-Civita connection. In general, however, the forms will be not closed and the connection compatible with the $SU(3)$ structure on $(M,g)$ will have non-vanishing torsion. The resulting geometries are classified according to the decomposition of the torsion $\tau$ into irreducible $SU(3)$ representations
\begin{equation}
\tau \in (\mathbf{1} + \mathbf{1})\oplus (\mathbf{8} + \mathbf{8})\oplus (\mathbf{6} + \mathbf{\bar{6}})\oplus (\mathbf{3} + \mathbf{\bar{3}})\oplus (\mathbf{3} + \mathbf{\bar{3}})
\end{equation}
which are referred to as the five torsion classes ${W}_{1}\,,{W}_{2}\,,{W}_{3}\,,{W}_{4}\,\text{and }{W}_{5}\,.$ In terms of these, $J$ and $\Omega$ can be expressed as
\begin{equation}\label{su3torsion}
	dJ=-\frac{3}{2}{\rm Im}(W_1\bar\Omega)+W_4\wedge J+W_3\; ,\qquad
	d\Omega=-W_1J\wedge J+W_2\wedge J+\bar W_5\wedge\Omega\; ,
\end{equation}
where the classes satisfy
\begin{equation}
	W_3\wedge J=W_3\wedge\Omega=W_2\wedge J\wedge J=0\; ,
\end{equation}
in order to fit the compatibility relations~\eqref{comp}. Here are some special types of $SU(3)$ structures which are characterized by the following conditions on the torsion classes~\footnote{In the mathematical literature half-flat manifolds are characterized by $\tau \in{W}_1^-\oplus{W}_2^-\oplus{W}_3$.}
\begin{center}
\begin{tabular}{ll}
   nearly Kaehler & $\tau \in {W}_1$ \\
   almost Kaehler & $\tau \in {W}_2$ \\
   Kaehler & $\tau \in{W}_5$ \\
   half-flat & $\tau \in{W}_1^+\oplus{W}_2^+\oplus{W}_3$.
  \end{tabular}
\end{center}
\vspace{0.2cm}
Furthermore, the Strominger system has the following characterisation in terms of the torsion classes,
\begin{equation}
\tau \in{W}_3\oplus{W}_4\oplus{W}_5\;, \quad 2W_4=W_5  \quad {\rm and} \quad W_4, W_5 \;{\rm are\; exact\; and\; real}.
\end{equation}

%%%
As previously stated, we are decomposing the ten dimensional space $M_{10}$ into 
\begin{equation}
M_3 \times M_1 \times X\; ,
\end{equation}
where $M_3$ is $2+1$-dimensional Minkowski space. The heterotic supergravity equations, in the absence of flux and with a constant dilaton, dictate that  $M_1 \times X$ is a manifold with $G_2$ holonomy. This is equivalent to the existence of a covariantly constant spinor $\eta$ on $M_1 \times X$. This spinor can be decomposed into two chiral spinors $\eta_\pm$ on $X$ by writing
\begin{equation}
\eta(x^m)=\frac{1}{\sqrt{2}}\left( \eta_+(x^m) + \eta_-(x^m)\right)\; .\label{eq_appendix_spinoransatz}
\end{equation}
This allows us to construct a real two-form and a complex three-form
\begin{equation}
	J_{uv}=- i\eta_+^\dagger\gamma_{uv}\eta_+\; ,\quad
	\Omega_{uvw}=\eta^\dagger_+\gamma_{uvw}\eta_-\; .
\end{equation}
It can be shown \cite{ChiossiSalamon, Hitchin}, given the appropriate flow of these tensors along $M_1$ as described by Eqs.~\eqref{HF1}, that those define an $SU(3)$ structure on the six-dimensional space which is half-flat.

%%%%%%%%%%%%
\section{The three coset spaces}\label{Data}
\setall
In this appendix we collect relevant information on the three coset spaces $G/H$ we focus on in this paper. This includes generators of the Lie-group, structure constants and some topological information such as Betti numbers. The generators, $T_A$ of the Lie-algebra of $G$ are split up as $\{T_A\}=\{K_a,H_i\}$, where $K_a$, $a=1,\ldots ,6$ denote the coset generators and $\{H_i\}$, $i=7,\ldots ,{\rm dim}(G)$ are the generators of the sub-group $H$. 

%%%%%%
\subsection{$SU(3)/U(1)^2$}\label{appendix_su3structureconstants}
Let us first recall the standard Gell-Mann matrices for the Lie algebra of $SU(3)$. 

\begin{eqnarray*}
&\lambda_1
=
-\frac{i}{2}\left(
\begin{array}{ccc}
 0 & 1 & 0 \\
 1 & 0 & 0 \\
 0 & 0 & 0
\end{array}
\right)
,\;
\lambda_2
=
\frac{1}{2}\left(
\begin{array}{ccc}
 0 & -1 & 0 \\
 1 & 0 & 0 \\
 0 & 0 & 0
\end{array}
\right)
,\;
\lambda_3
=
-\frac{i}{2}\left(
\begin{array}{ccc}
 1 & 0 & 0 \\
 0 & -1 & 0 \\
 0 & 0 & 0
\end{array}
\right)
,
\\
&\lambda_4
=
-\frac{i}{2}\left(
\begin{array}{ccc}
 0 & 0 & 1 \\
 0 & 0 & 0 \\
 1 & 0 & 0
\end{array}
\right)
,\;
\lambda_5
=\frac{1}{2}\left(
\begin{array}{ccc}
0 & 0 & -1 \\
 0 & 0 & 0 \\
 1 & 0 & 0

\end{array}
\right)
,\;
\lambda_6
=-\frac{i}{2}\left(
\begin{array}{ccc}
 0 & 0 & 0 \\
 0 & 0 & 1 \\
 0 & 1 & 0
\end{array}
\right)
,
\\
&
\lambda_7
=\frac{1}{2}\left(
\begin{array}{ccc}
  0 & 0 & 0 \\
 0 & 0 & -1 \\
 0 & 1 & 0
\end{array}
\right)
,\;
\lambda_8
=-\frac{i}{2\sqrt{3}}
\left(
\begin{array}{ccc}
 1 & 0 & 0 \\
 0 & 1 & 0 \\
 0 & 0 & -2
\end{array}
\right)
\;.
\end{eqnarray*}
Our generators are given by a re-labelled version of the Gell-Mann matrices defined as follows. 
\begin{eqnarray}
&K_1=\lambda_1\;,\quad K_2=\lambda_2\;,\quad K_3=\lambda_4\;,\quad K_4=\lambda_5\;,\\
&K_5=\lambda_6\;,\quad K_6=\lambda_7\;,\quad H_7=\lambda_3\;,\quad H_8=\lambda_8\;.
\end{eqnarray}
The geometry of the homogeneous space $SU(3)/U(1)\times U(1)$ is entirely determined by the structure constants which in our basis $\{K_a, H_i\}$ are given by
\begin{eqnarray}\nonumber
&&f_{12}^{\phantom{12}7}=1\\
&&f_{13}^{\phantom{13}6}=
-f_{14}^{\phantom{14}5}=
f_{23}^{\phantom{23}5}=
f_{24}^{\phantom{24}6}=
f_{73}^{\phantom{73}4}=
-f_{75}^{\phantom{75}6}=1/2\\\nonumber
&&f_{34}^{\phantom{34}8}=
f_{56}^{\phantom{56}8}=\sqrt{3}/2
\;.
\end{eqnarray}

%%%
The non-vanishing Betti numbers of this coset are given by~\cite{MuellerHoissen:1987cq}
\begin{equation}
	b_0=1\;, \quad \quad b_2=2\;,\quad \quad b_4=2\;,\quad \quad b_6=1\; ,
\end{equation}
which leads to the Euler number
\begin{equation}
	\chi=6 \;.
\end{equation}

%%%%%%
\subsection{$Sp(2)/SU(2)\times U(1)$}\label{appendix_sp2structureconstants}

We choose as generators for $Sp(2)$
\begin{eqnarray*}
&
K_1
=\frac{1}{\sqrt{2}}
\left(
\begin{array}{cccc}
 0 & 0 & 1 & 0 \\
 0 & 0 & 0 & 1 \\
 -1 & 0 & 0 & 0 \\
 0 & -1 & 0 & 0
\end{array}
\right),\;
K_2=\frac{i}{\sqrt{2}}\left(
\begin{array}{cccc}
 0 & 0 & 0 & 1 \\
 0 & 0 & 1 & 0 \\
 0 & 1 & 0 & 0 \\
 1 & 0 & 0 & 0
\end{array}
\right) 
,\\
&
K_3= \left(
\begin{array}{cccc}
 i & 0 & 0 & 0 \\
 0 & -i & 0 & 0 \\
 0 & 0 & 0 & 0 \\
 0 & 0 & 0 & 0
\end{array}
\right)
,
K_4
=
\left(
\begin{array}{cccc}
 0 & 1 & 0 & 0 \\
 -1 & 0 & 0 & 0 \\
 0 & 0 & 0 & 0 \\
 0 & 0 & 0 & 0
\end{array}
\right)
,
K_5
=\frac{1}{\sqrt{2}}\left(
\begin{array}{cccc}
 0 & 0 & 0 & 1 \\
 0 & 0 & -1 & 0 \\
 0 & 1 & 0 & 0 \\
 -1 & 0 & 0 & 0
\end{array}
\right)
\\
&
K_6=\frac{i}{\sqrt{2}}\left(
\begin{array}{cccc}
 0 & 0 & -1 & 0 \\
 0 & 0 & 0 & 1 \\
 -1 & 0 & 0 & 0 \\
 0 & 1 & 0 & 0
\end{array}
\right)
,\;
H_7
=
\left(
\begin{array}{cccc}
 0 & 0 & 0 & 0 \\
 0 & 0 & 0 & 0 \\
 0 & 0 & i & 0 \\
 0 & 0 & 0 & -i
\end{array}
\right)
,\\
&
H_8
=
\left(
\begin{array}{cccc}
 0 & 0 & 0 & 0 \\
 0 & 0 & 0 & 0 \\
 0 & 0 & 0 & -1 \\
 0 & 0 & 1 & 0
\end{array}
\right)
,\;
H_9=\left(
\begin{array}{cccc}
 0 & 0 & 0 & 0 \\
 0 & 0 & 0 & 0 \\
 0 & 0 & 0 & -i \\
 0 & 0 & -i & 0
\end{array}
\right)
,\;
H_{10}=\left(
\begin{array}{cccc}
 0 & i & 0 & 0 \\
 i & 0 & 0 & 0 \\
 0 & 0 & 0 & 0 \\
 0 & 0 & 0 & 0
\end{array}
\right)
\end{eqnarray*}
The respective decomposition of the $Sp(2)$ Lie algebra corresponds to the non-maximal embedding of $SU(2)\times U(1)$ into $Sp(2)$. The homogeneous space $Sp(2)/SU(2)\times U(1)$ is entirely determined by the associated structure constants
\begin{eqnarray}\nonumber
&&f_{13}^{\phantom{13}6}=
-f_{14}^{\phantom{14}5}=
f_{23}^{\phantom{23}5}=
f_{24}^{\phantom{24}6}=1\\
&&f_{71}^{\phantom{71}6}=
-f_{72}^{\phantom{72}5}=
f_{81}^{\phantom{81}5}=
f_{82}^{\phantom{82}6}=
f_{91}^{\phantom{91}2}=
-f_{95}^{\phantom{95}6}=
f_{10\;1}^{\phantom{10\;1}2}=
f_{10\;5}^{\phantom{10\;5}6}=1\\\nonumber
&&f_{78}^{\phantom{78}9}=
f_{10\;3}^{\phantom{10\;3}4}=2
\;.
\end{eqnarray}

%%%
The non-vanishing Betti numbers are~\cite{MuellerHoissen:1987cq}
\begin{equation}
	b_0=1\;, \quad \quad b_2=1\;,\quad \quad b_4=1\;,\quad \quad b_6=1\; ,
\end{equation}
resulting in the Euler number
\begin{equation}
	\chi=4 \;.
\end{equation}

%%%%%%
\subsection{$G_2/SU(3)$}\label{appendix_g2structureconstants}

Our chosen generators for $G_2$ are
\begin{eqnarray*}
&
K_1=\frac{1}{\sqrt{3}}\left(
\begin{array}{ccccccc}
 0 & 2 & 0 & 0 & 0 & 0 & 0 \\
 -2 & 0 & 0 & 0 & 0 & 0 & 0 \\
 0 & 0 & 0 & 0 & 0 & 0 & 0 \\
 0 & 0 & 0 & 0 & 0 & 0 & 1 \\
 0 & 0 & 0 & 0 & 0 & 1 & 0 \\
 0 & 0 & 0 & 0 & -1 & 0 & 0 \\
 0 & 0 & 0 & -1 & 0 & 0 & 0
\end{array}
\right)
,\;
K_2=\frac{1}{\sqrt{3}}\left(
\begin{array}{ccccccc}
 0 & 0 & 2 & 0 & 0 & 0 & 0 \\
 0 & 0 & 0 & 0 & 0 & 0 & 0 \\
 -2 & 0 & 0 & 0 & 0 & 0 & 0 \\
 0 & 0 & 0 & 0 & 0 & 1 & 0 \\
 0 & 0 & 0 & 0 & 0 & 0 & -1 \\
 0 & 0 & 0 & -1 & 0 & 0 & 0 \\
 0 & 0 & 0 & 0 & 1 & 0 & 0
\end{array}
\right),
\\
&
K_3=\frac{1}{\sqrt{3}}\left(
\begin{array}{ccccccc}
 0 & 0 & 0 & 0 & -2 & 0 & 0 \\
 0 & 0 & 0 & 0 & 0 & 1 & 0 \\
 0 & 0 & 0 & 0 & 0 & 0 & -1 \\
 0 & 0 & 0 & 0 & 0 & 0 & 0 \\
 2 & 0 & 0 & 0 & 0 & 0 & 0 \\
 0 & -1 & 0 & 0 & 0 & 0 & 0 \\
 0 & 0 & 1 & 0 & 0 & 0 & 0
\end{array}
\right)
,\;
K_4=\frac{1}{\sqrt{3}}\left(
\begin{array}{ccccccc}
 0 & 0 & 0 & -2 & 0 & 0 & 0 \\
 0 & 0 & 0 & 0 & 0 & 0 & 1 \\
 0 & 0 & 0 & 0 & 0 & 1 & 0 \\
 2 & 0 & 0 & 0 & 0 & 0 & 0 \\
 0 & 0 & 0 & 0 & 0 & 0 & 0 \\
 0 & 0 & -1& 0 & 0 & 0 & 0 \\
 0 & -1 & 0 & 0 & 0 & 0 & 0
\end{array}
\right)
,
\end{eqnarray*}
\begin{eqnarray*}
&
K_5=\frac{1}{\sqrt{3}}\left(
\begin{array}{ccccccc}
 0 & 0 & 0 & 0 & 0 & 0 & 2 \\
 0 & 0 & 0 & 1 & 0 & 0 & 0 \\
 0 & 0 & 0 & 0 & -1 & 0 & 0 \\
 0 & -1 & 0 & 0 & 0 & 0 & 0 \\
 0 & 0 & 1 & 0 & 0 & 0 & 0 \\
 0 & 0 & 0 & 0 & 0 & 0 & 0 \\
 -2 & 0 & 0 & 0 & 0 & 0 & 0
\end{array}
\right)
,\;
K_6=\frac{1}{\sqrt{3}}\left(
\begin{array}{ccccccc}
 0 & 0 & 0 & 0 & 0 & 2 & 0 \\
 0 & 0 & 0 & 0 & 1 & 0 & 0 \\
 0 & 0 & 0 & 1 & 0 & 0 & 0 \\
 0 & 0 & -1 & 0 & 0 & 0 & 0 \\
 0 & -1 & 0 & 0 & 0 & 0 & 0 \\
 -2 & 0 & 0 & 0 & 0 & 0 & 0 \\
 0 & 0 & 0 & 0 & 0 & 0 & 0
\end{array}
\right)
,
\end{eqnarray*}
\begin{eqnarray*}
&
H_7=\left(
\begin{array}{ccccccc}
 0 & 0 & 0 & 0 & 0 & 0 & 0 \\
 0 & 0 & 0 & 0 & 0 & 0 & 0 \\
 0 & 0 & 0 & 0 & 0 & 0 & 0 \\
 0 & 0 & 0 & 0 & 0 & 0 & -1 \\
 0 & 0 & 0 & 0 & 0 & 1 & 0 \\
 0 & 0 & 0 & 0 & -1 & 0 & 0 \\
 0 & 0 & 0 & 1 & 0 & 0 & 0
\end{array}
\right)
,\;
H_8=\left(
\begin{array}{ccccccc}
 0 & 0 & 0 & 0 & 0 & 0 & 0 \\
 0 & 0 & 0 & 0 & 0 & 0 & 0 \\
 0 & 0 & 0 & 0 & 0 & 0 & 0 \\
 0 & 0 & 0 & 0 & 0 & -1 & 0 \\
 0 & 0 & 0 & 0 & 0 & 0 & -1 \\
 0 & 0 & 0 & 1 & 0 & 0 & 0 \\
 0 & 0 & 0 & 0 & 1 & 0 & 0
\end{array}
\right)
,
\end{eqnarray*}
\begin{eqnarray*}
&
H_9=\left(
\begin{array}{ccccccc}
 0 & 0 & 0 & 0 & 0 & 0 & 0 \\
 0 & 0 & 0 & 0 & 0 & 0 & 0 \\
 0 & 0 & 0 & 0 & 0 & 0 & 0 \\
 0 & 0 & 0 & 0 & -1 & 0 & 0 \\
 0 & 0 & 0 & 1 & 0 & 0 & 0 \\
 0 & 0 & 0 & 0 & 0 & 0 & 1 \\
 0 & 0 & 0 & 0 & 0 & -1 & 0
\end{array}
\right)
,\;
H_{10}=\left(
\begin{array}{ccccccc}
 0 & 0 & 0 & 0 & 0 & 0 & 0 \\
 0 & 0 & 0 & 0 & 0 & -1 & 0 \\
 0 & 0 & 0 & 0 & 0 & 0 & -1 \\
 0 & 0 & 0 & 0 & 0 & 0 & 0 \\
 0 & 0 & 0 & 0 & 0 & 0 & 0 \\
 0 & 1 & 0 & 0 & 0 & 0 & 0 \\
 0 & 0 & 1 & 0 & 0 & 0 & 0
\end{array}
\right)
,
\end{eqnarray*}
\begin{eqnarray*}
&
H_{11}=\left(
\begin{array}{ccccccc}
 0 & 0 & 0 & 0 & 0 & 0 & 0 \\
 0 & 0 & 0 & 0 & 0 & 0 & 1 \\
 0 & 0 & 0 & 0 & 0 & -1 & 0 \\
 0 & 0 & 0 & 0 & 0 & 0 & 0 \\
 0 & 0 & 0 & 0 & 0 & 0 & 0 \\
 0 & 0 & 1 & 0 & 0 & 0 & 0 \\
 0 & -1 & 0 & 0 & 0 & 0 & 0
\end{array}
\right)
,\;
H_{12}=\left(
\begin{array}{ccccccc}
 0 & 0 & 0 & 0 & 0 & 0 & 0 \\
 0 & 0 & 0 & 1 & 0 & 0 & 0 \\
 0 & 0 & 0 & 0 & 1 & 0 & 0 \\
 0 & -1 & 0 & 0 & 0 & 0 & 0 \\
 0 & 0 & -1 & 0 & 0 & 0 & 0 \\
 0 & 0 & 0 & 0 & 0 & 0 & 0 \\
 0 & 0 & 0 & 0 & 0 & 0 & 0
\end{array}
\right)
,
\end{eqnarray*}
\begin{eqnarray*}
&
H_{13}=\left(
\begin{array}{ccccccc}
 0 & 0 & 0 & 0 & 0 & 0 & 0 \\
 0 & 0 & 0 & 0 & -1 & 0 & 0 \\
 0 & 0 & 0 & 1 & 0 & 0 & 0 \\
 0 & 0 & -1 & 0 & 0 & 0 & 0 \\
 0 & 1 & 0 & 0 & 0 & 0 & 0 \\
 0 & 0 & 0 & 0 & 0 & 0 & 0 \\
 0 & 0 & 0 & 0 & 0 & 0 & 0
\end{array}
\right)
,\;
H_{14}=\frac{1}{\sqrt{3}}\left(
\begin{array}{ccccccc}
 0 & 0 & 0 & 0 & 0 & 0 & 0 \\
 0 & 0 & -2 & 0 & 0 & 0 & 0 \\
 0 & 2 & 0 & 0 & 0 & 0 & 0 \\
 0 & 0 & 0 & 0 & 1 & 0 & 0 \\
 0 & 0 & 0 & -1 & 0 & 0 & 0 \\
 0 & 0 & 0 & 0 & 0 & 0 & 1 \\
 0 & 0 & 0 & 0 & 0 & -1 & 0
\end{array}
\right)
\;.
\end{eqnarray*}
The structure constant associated to this choice of generators are
\begin{eqnarray}\nonumber
&&f_{7\; 10}^{\phantom{7\; 10}13}=
-f_{7\; 11}^{\phantom{7\; 11}12}=
f_{73}^{\phantom{73}6}=
-f_{74}^{\phantom{74}5}=1\\\nonumber&&
f_{8\; 10}^{\phantom{8\; 10}12}=
f_{8\; 11}^{\phantom{8\; 11}13}=
-f_{83}^{\phantom{83}5}=
-f_{84}^{\phantom{84}6}=
f_{9\; 10}^{\phantom{9\; 10}11}=
-f_{9\; 12}^{\phantom{9\; 12}13}=
-f_{93}^{\phantom{93}4}=
f_{95}^{\phantom{95}6}=1\\\nonumber&&
f_{10\; 1}^{\phantom{10\; 1}6}=
f_{10\; 2}^{\phantom{10\; 2}5}=
-f_{11\; 1}^{\phantom{11\; 1}5}=
f_{11\; 2}^{\phantom{11\; 2}6}=
f_{12\; 1}^{\phantom{12\; 1}4}=
f_{12\; 2}^{\phantom{12\; 2}3}=
-f_{13\; 1}^{\phantom{13\; 1}3}=
f_{13\; 2}^{\phantom{13\; 2}4}=1\\
&&f_{10\; 11}^{\phantom{10\; 11}14}=
f_{12\; 13}^{\phantom{12\; 13}14}=\sqrt{3},\quad\quad
f_{78}^{\phantom{78}9}=2\\\nonumber&&
f_{14\; 1}^{\phantom{14\; 1}2}=
f_{13}^{\phantom{13}6}=
f_{14}^{\phantom{14}5}=
-f_{23}^{\phantom{23}5}=
f_{24}^{\phantom{24}6}=2/\sqrt{3}\\\nonumber&&
f_{14\; 3}^{\phantom{14\; 3}4}=
f_{14\; 5}^{\phantom{14\; 5}6}=1/\sqrt{3}
\;.
\end{eqnarray}

The non-vanishing Betti numbers are~\cite{MuellerHoissen:1987cq}
\begin{equation}
	b_0=1\;, \quad \quad  b_6=1\;.
\end{equation}
which gives the Euler number
\begin{equation}
	\chi=2 \;.
\end{equation}

%%%%%%%%%%%%%%%%%%%%%%%%%%%%%%%%%%%%%%%%%%%%%%%%%%%%%%%%%%%%%%%%%%%%%%%%%%%%%%%%%%%%%%%%%%%%%%%%%%%%%%%%%%%%%%%%%%%%%%%%%%%%%%%%%%%%%%%%%%%%%%%%

\end{document}